\documentclass[conference]{IEEEtran}
\IEEEoverridecommandlockouts
\usepackage{cite}  
\usepackage{tabu}                      
\usepackage{booktabs}                  
\usepackage{lipsum}                    
\usepackage{mwe}                       
\usepackage[T1]{fontenc}
\usepackage{epsfig}
\usepackage{subcaption}
\usepackage{calc}

\usepackage{graphicx}
\usepackage{booktabs}
\usepackage{soul}

\usepackage{cite}
\usepackage{amsmath,amssymb,amsfonts}
\usepackage{subcaption}
\usepackage{algorithm}
\usepackage{algpseudocode}
\usepackage{textcomp}
\usepackage{xcolor}
\usepackage{multicol}
\usepackage{multirow}
\usepackage{makecell}
\usepackage{hyperref}
\usepackage{pgf}
\usepackage{pgfplots}
\pgfplotsset{compat=1.18}
\usepgfplotslibrary{statistics}

\definecolor{sblue}{HTML}{4C72B0}
\definecolor{sorange}{HTML}{DD8452}
\definecolor{sgreen}{HTML}{55A868}

\usepackage{threeparttable}

\usepackage{listings}

\definecolor{promptbg}{RGB}{244,250,244}
\definecolor{promptborder}{RGB}{170,190,170}
\definecolor{prompttext}{RGB}{35,45,35}
\definecolor{promptaccent}{RGB}{70,120,80}
\definecolor{mintbackground}{HTML}{F0FFF4}
\definecolor{mintborder}{HTML}{C6F6D5}

\lstdefinestyle{promptstyle}{
    basicstyle=\ttfamily\normalsize\color{black!95},
    breaklines=true,
    breakatwhitespace=true,
    keepspaces=true,
    columns=flexible,
    frame=single,
    framerule=0.6pt,
    rulecolor=\color{mintborder},
    backgroundcolor=\color{mintbackground!20},
    xleftmargin=8pt,
    xrightmargin=8pt,
    aboveskip=12pt,
    belowskip=8pt,
    showstringspaces=false,
    upquote=true,
    literate={'}{\textquotesingle}1,
}

\usepackage{url}
\usepackage{adjustbox}
\usepackage{rotating}
\def\BibTeX{{\rm B\kern-.05em{\sc i\kern-.025em b}\kern-.08em
    T\kern-.1667em\lower.7ex\hbox{E}\kern-.125emX}}
\begin{document}
\title{TexSketch: Bringing Texture-Aware Colorization to Sketches}
\author{
\IEEEauthorblockN{Taraash Mittal\IEEEauthorrefmark{1}, Gaurav Rai\IEEEauthorrefmark{1}, 
Ojaswa Sharma}

\IEEEauthorblockA{Graphics Research Group, IIIT Delhi}
\{taraash23552, gauravr, ojaswa\}@iiitd.ac.in\\
\IEEEauthorrefmark{1} denotes equal contribution}

\maketitle
\begin{abstract}
Reference-based sketch colorization methods rely on large paired datasets that preserve both the structural and stylistic characteristics of hand-drawn artwork. However, existing datasets are limited in scale, expensive to annotate, and bound to fixed, often inconsistent artistic style biases that propagate to downstream models and limit cross-domain generalization. We present TexSketch, a controllable procedural framework for generating colored-sketch datasets with programmable artistic styles via geometric analysis and shader-driven stylization. Our fully automatic pipeline integrates region extraction, semantic color prediction, and shader-based rendering. By defining artistic appearance procedurally rather than inheriting it from a static corpus, TexSketch enables scalable dataset generation without manual annotation or artist supervision. Human studies demonstrate that TexSketch generates perceptually plausible colored sketches with high stylistic diversity, providing a controllable, scalable source of synthetic supervision for sketch colorization.
\end{abstract}

\section{Introduction}
Sketch colorization is a well-known research field due to its applications in digital art creation, animation production, and artistic content generation. Modern learning-based approaches~\cite{zhang2025animecolor, yan2025colorizediffusion}, particularly diffusion and transformer-based models, rely heavily on large-scale paired datasets of sketches and corresponding colored references to achieve robust appearance transfer and stylistic consistency. However, existing colored sketch datasets remain limited in scale, stylistic diversity, and annotation quality. Existing collections are heavily concentrated around anime and manga illustrations, and this bias is inherited by downstream colorization models~\cite{qiu2025mangadit}. As a result, models trained on these data often generalize poorly to watercolor, colored-pencil, concept art, and other non-anime domains. Constructing such datasets manually is both labor-intensive and artistically inconsistent, as hand-colored sketches require significant human expertise and stylistic control. Consequently, the lack of large-scale, high-quality paired sketch–color datasets has become a major bottleneck for training and evaluating sketch colorization systems.

To address these limitations, we introduce TexSketch, a controllable procedural framework for generating synthetic colored sketches. Unlike existing approaches that learn artistic styles from fixed datasets, TexSketch explicitly defines artistic appearance through region-based semantic color assignment and Open Shading Language (OSL)-based programmable rendering \cite{gritz2010open}. Given an input sketch, our framework first segments the image into semantically coherent regions using the SegmentAnything Model~\cite{ravi2025sam}. Further, it predicts perceptually plausible colors for each region using Qwen-guided vision-language reasoning \cite{Qwen-VL}, computes geometric descriptors, and applies programmable stylization to synthesize textured colored sketches. This procedural formulation offers explicit, fine-grained control over artistic style and surface texture, enabling the generation of diverse visual domains from a single underlying sketch line-drawing.

Our primary contributions are summarized as follows:
\begin{itemize}
\item We present TexSketch, a scalable, supervision-free framework for controllable colored-sketch synthesis that significantly reduces reliance on large, domain-specific datasets.
\item We propose a novel pipeline that pairs vision-language semantic color planning with geometric descriptors to guide shader-driven rendering.
\item We demonstrate explicit control over non-photorealistic textures, enabling the procedural synthesis of diverse artistic mediums (e.g., watercolor, cel-shading) that break free from anime-centric biases.
\item We validate our method through human perceptual studies, demonstrating that it produces highly plausible, style-consistent colorization that is effective for synthetic supervision.
\end{itemize}

\section{Related Work}

\subsection{Sketch Datasets}
Large-scale sketch datasets have been widely explored for sketch recognition, retrieval, segmentation, and sketch-based synthesis. Early benchmarks such as TU-Berlin~\cite{eitz2012humans} and QuickDraw~\cite{ha2017neural} provide large collections of stroke-based sketches across diverse object categories, while subsequent datasets including SketchSeg150K~\cite{qi2019sketchsegnet+}, SketchFix-160~\cite{sarvadevabhatla2017object}, Sketchy~\cite{sangkloy2016sketchy}, and SketchyScene~\cite{zou2018sketchyscene} introduce richer structural and scene-level annotations. Despite their scale and structural diversity, these benchmarks are primarily optimized for recognition and retrieval tasks, and they generally lack the paired, aligned colored counterparts required for supervised sketch colorization. 
Existing paired sketch-color datasets~\cite{xian2018texturegan, sangkloy2017scribbler} are largely derived from anime and manga illustrations, leading to a strong bias toward flat, cel-shaded aesthetics. Consequently, models trained on such data often generalize inaccurately to textured sketches, artistic line art, and non-anime drawing styles~\cite{danbooru2021}. These limitations highlight the need for scalable, stylistically diverse colored sketch datasets to enable more generalizable sketch colorization models.

\subsection{Sketch Colorization}
Sketch colorization has evolved from interactive editing systems to modern learning-based approaches. Early methods, such as Zhang et al.~\cite{zhang2018two}, propose a method that generates coarse color predictions and then refines them to improve local consistency. Subsequently, to address the limitations of handling incomplete line structures and ambiguous topology, geometry-driven approaches such as Delaunay Painting~\cite{parakkat2022delaunay} introduced triangulation-based segmentation for raster sketches. For vector graphics, Scrivener et al.~\cite{scrivener2024winding} exploited topological winding number representations for robust region filling, while KISSColor~\cite{dong2025kisscolor} and Ciallo~\cite{ciao2024ciallo} improved interactive vector colorization and real-time GPU-accelerated rendering. Reference-based approaches provide external guidance by directly transferring colors from exemplar images. Lee et al.~\cite{lee2020reference} introduce dense semantic correspondence to achieve structure-preserving color transfer, while recent diffusion-based methods, including ColorizeDiffusion v2~\cite{yan2025colorizediffusion} and MagicColor~\cite{zhang2025magiccolor}, improve synthesis quality and multi-instance consistency through diffusion-driven color propagation. Furthermore, SketchDeco~\cite{utintu2026sketchdeco} uses predefined color palettes and spatial masks to perform precise latent-space blending, adding reference colors to the sketch. Despite these advances, existing methods~\cite{yan2025image, zhang2025animecolor, qiu2025mangadit, yan2026towards} are predominantly constrained on anime and manga datasets, resulting in flat, cel-shaded outputs with limited texture diversity. Consequently, they often generalize inaccurately to textured sketches, artistic line art, and traditional hand-drawn styles. In contrast, our work focuses on procedural dataset generation for textured sketch colorization, using geometry-aware shader synthesis to create stylistically diverse colored sketches for training more generalizable colorization models.

\section{Methodology}
\begin{figure*}[t]
    \centering
    \includegraphics[width=0.95\linewidth]{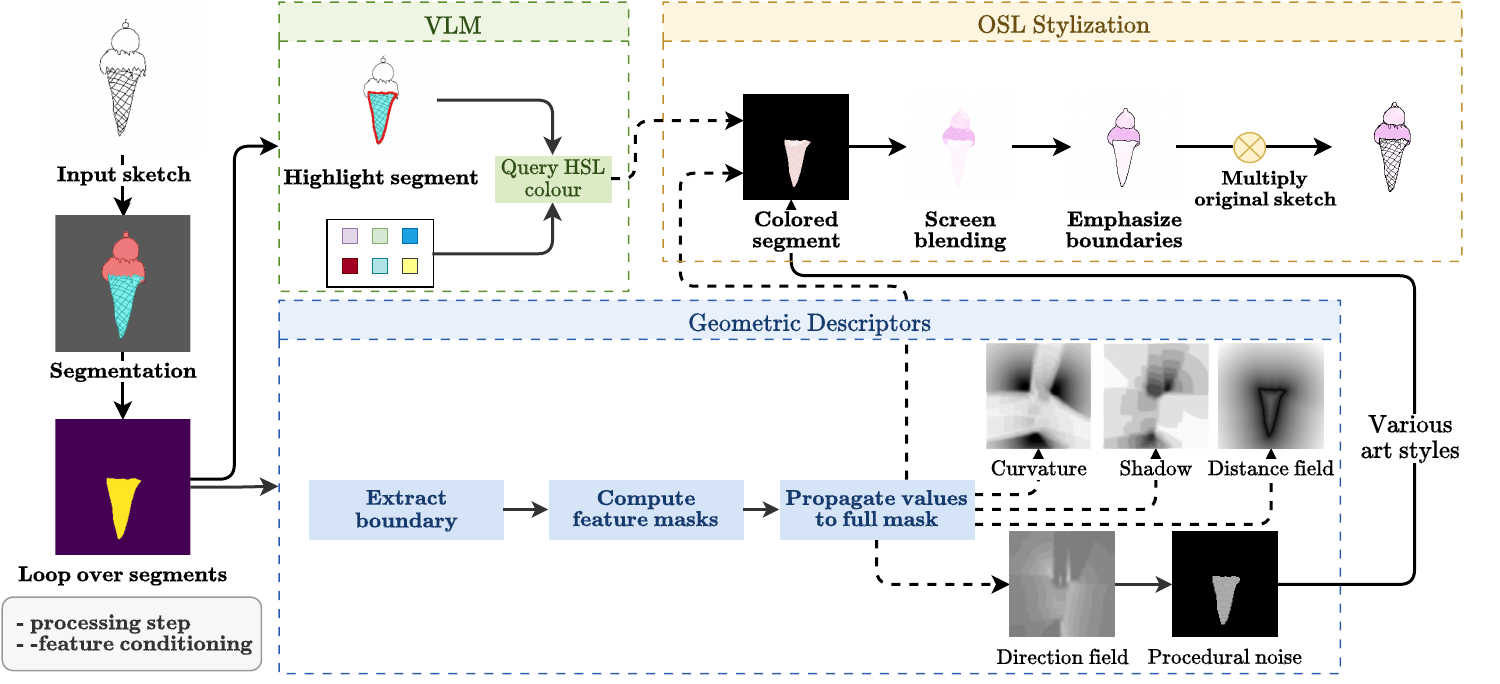}
    \caption{Overview of the proposed TexSketch pipeline. The input sketch is first partitioned into semantically coherent regions, which are assigned semantic base colors using a vision-language model. Geometry-aware descriptor fields are then extracted for each region and used to guide procedural OSL shaders, producing the final colored sketch.}
    \label{fig:pipeline}
\end{figure*}

Given a grayscale input sketch $I \in \mathbb{R}^{H \times W}$, our pipeline sequentially performs region extraction, semantic color assignment, geometric feature extraction, procedural stylization, and compositing, as illustrated in Fig.~\ref{fig:pipeline}. By systematically varying procedural rendering parameters and semantic color strategies, our framework can synthesize a single sketch into distinct art styles.

\subsection{Sketch Segmentation}
We begin by preprocessing the input sketch $I$ with contrast normalization and bilateral filtering to facilitate its partitioning into semantically coherent regions. This preprocessing stage suppresses high-frequency noise and scanning artifacts without degrading anti-aliased contours, thereby retaining fine boundary information for precise semantic segmentation.
The preprocessed grayscale sketch is subsequently converted into a three-channel RGB image and passed to the Segment Anything Model (SAM)~\cite{ravi2025sam} operating in automatic mask generation mode. The resulting candidate masks are filtered based on their pixel area and image-border adjacency, and are subsequently smoothed to remove aliasing artifacts. This process eliminates background-dominated regions while retaining a stable set of closed region proposals, ${R_i}_{i=1}^N$.

\subsection{Semantic Color Assignment}
For each extracted region $R_i$, we predict a semantically meaningful color represented in the $(H, S, L) \in [0, 360] \times [0, 1] \times [0, 1]$ (Hue, Saturation, Lightness) color space. We use a vision-language model (VLM)~\cite{Qwen-VL} to analyze the localized mask of $R_i$ within the global context of the input sketch $I$. To ensure global color consistency, the model is conditioned on the current region mask, the input sketch, and a structured history of the colors assigned to previously processed regions. This context-aware conditioning encourages semantically coherent color assignments across related regions (e.g., skin, hair, and clothing) and preserves perceptually distinct color transitions along adjacent boundaries. We constrain the VLM to produce a structured JSON output containing an HSL color triplet for the target region along with the corresponding selection rationale. The predicted HSL triplet serves as the base albedo color for the subsequent procedural rendering stage. By decoupling semantic color prediction from spatial texture synthesis, TexSketch enables independent and fine-grained control over region semantics, local shading, and the overall artistic style.

\subsection{Geometric Feature Extraction}
\label{sec:geometry}
To synthesize effective hand-drawn artwork, our procedural shaders require local geometric cues that mimic the structural characteristics an artist implicitly follows when shading and stylizing a sketch. For each segmented region, we compute a dense set of boundary-derived geometric descriptor fields, including the \textit{distance-to-edge}, \textit{boundary curvature}, \textit{directional shadow}, and \textit{direction field}. These descriptors provide spatial information that guides texture synthesis, stroke orientation, and shading within each region. Since these geometric descriptors are intrinsically defined only along the region boundary, they must be propagated to the interior to provide dense guidance for procedural rendering. For each pixel $\mathbf{p}$ within a region, we identify its nearest boundary pixel, denoted by $\mathbf{e}{\mathbf{p}}=\mathrm{nearest}(\mathbf{p})$, using the Jump Flooding Algorithm (JFA)~\cite{rong2006jump}. The boundary-defined descriptor value, $f(\mathbf{e}{\mathbf{p}})$, is then propagated to the interior and weighted by a radial distance falloff function:
\begin{equation}
    w(d(\mathbf{p})) = \exp\left(-\frac{d(\mathbf{p})}{\sigma}\right),
\end{equation}
where $d(\mathbf{p}) = \|\mathbf{p} - \mathbf{e}_{\mathbf{p}}\|_2$ represents the Euclidean distance to the closest boundary, and $\sigma$ is a structural decay hyperparameter. Finally, a small box blur is applied to the dense descriptor fields to overcome Voronoi boundary artifacts arising from the discrete nearest-neighbor propagation, ensuring spatially smooth and continuous guidance fields.

\subsubsection{Distance to Edge}
The distance-to-edge descriptor quantifies the proximity of each pixel to the region boundary. Let $E$ denote the set of coordinate pixels forming boundary contour of the segmented region. The Euclidean distance field is formally defined as:
\begin{equation}
    d(\mathbf{p}) = \min_{\mathbf{e}\in E} \|\mathbf{p} - \mathbf{e}\|_2.
\end{equation}
We generate two normalized 8-bit representations of the distance field, each linearly mapped to the range $[0,255]$:

\paragraph{Global Distance Field:} To preserve a consistent distance scale across all regions, the distance values are normalized by the maximum possible distance within the image canvas, $d_{\max}$:
\begin{equation}
    M_{\text{global}}(\mathbf{p}) = \left(\frac{d(\mathbf{p})}{d_{\max}}\right)^{\gamma},
\end{equation}
where $\gamma = 0.5$ acts as a power-law contrast adjustment to emphasize gradients near boundaries.
\paragraph{Interior Distance Field:} To preserve the relative depth structure within each boundary projection, the distance values are normalized by the maximum interior distance measured along the corresponding projection line. Let $L_{\mathbf{e}} = \max_{\{\mathbf{p} \mid \mathrm{nearest}(\mathbf{p})=\mathbf{e}\}} d(\mathbf{p})$ be the maximum distance mapped to a boundary pixel $\mathbf{e}$. The interior mask is defined as:
\begin{equation}
    M_{\text{interior}}(\mathbf{p}) = \left(\frac{d(\mathbf{p})}{L_{\mathrm{nearest}(\mathbf{p})}}\right)^{\gamma}.
\end{equation}

The final distance mask then becomes

\[
M_{\text{dist}}(\mathbf{p}) =
\begin{cases}
M_{\text{interior}}, & \text{if } \mathbf{p} \text{ is interior},\\
M_{\text{global}}, & \text{otherwise}.
\end{cases}
\]

\subsubsection{Boundary Curvature}
Local boundary curvature provides an important geometric cue for procedural shading, as artists often accumulate ink or pigments in highly curved crevices. We estimate local boundary curvature using a ring-intersection method. For each boundary pixel $\mathbf{e}\in E$, a circular neighborhood of radius $R$ centered at $\mathbf{e}$ is sampled to identify its intersection points with the boundary contour $E$. These intersections are partitioned into two spatial branches using $k$-means clustering ($k=2$), yielding cluster centroids $\mathbf{a}$ and $\mathbf{b}$. We construct the normalized directional arms as:
\begin{equation}
    \mathbf{v}_A = \frac{\mathbf{a}-\mathbf{e}}{\|\mathbf{a}-\mathbf{e}\|_2}, \quad \mathbf{v}_B = \frac{\mathbf{b}-\mathbf{e}}{\|\mathbf{b}-\mathbf{e}\|_2}.
\end{equation}
The interior angle between these arms is given by:
\begin{equation}
    \theta = \arccos\big(\mathrm{clamp}(\mathbf{v}_A \cdot \mathbf{v}_B, -1.0, 1.0)\big).
\end{equation}
Using a local Principal Component Analysis (PCA) framework~\cite{10.1145/3447755} over a localized neighborhood around $\mathbf{e}$, we compute a smoothed tangent vector $\mathbf{t}$ and a corresponding curvature normal vector $\mathbf{n}$ (directed via projection toward the local centroid) to act as downstream shading coordinates. Subsequently, the boundary-defined curvature values are propagated throughout the region interior using the JFA-derived nearest-edge map and are attenuated according to the exponential distance falloff function. The resulting curvature value at each interior pixel defines the filled curvature mask $M_\text{curv}$, which is subsequently smoothed using a box filter to suppress residual high-frequency Voronoi artifacts.

\subsubsection{Shadow Proxy}
To introduce depth cues and volumetric variation into the synthesized training pairs, we simulate directional illumination using a randomly oriented global light source. The illumination direction is defined by an orientation angle $\alpha$, sampled uniformly from $\mathcal{U}(0,2\pi)$, from which the corresponding two-dimensional unit light vector is computed as $\mathbf{L}=(\cos\alpha,\sin\alpha)$.
At each boundary pixel $\mathbf{e} \in E$, we retrieve the PCA-smoothed tangent vector $\mathbf{t}$ and rotate it by $90^\circ$ counter-clockwise to establish a localized boundary normal $\mathbf{n}$. The boundary shadow value $s_{\mathbf{e}}$ is evaluated via the absolute lambertian proxy:
\begin{equation}
    s_{\mathbf{e}} = |\mathbf{n} \cdot \mathbf{L}| \in [0,1].
\end{equation}
The boundary-defined shadow intensities are propagated throughout the region interior using the JFA-derived nearest-edge map and are attenuated according to the exponential distance falloff function with $\sigma = 60.0$. The resulting shadow value at each interior pixel $\mathbf{p}$ is given by:
\begin{equation}
    M_{\text{shadow}}(\mathbf{p}) = s_{\mathrm{nearest}(\mathbf{p})} \cdot \exp\left(-\frac{d(\mathbf{p})}{\sigma}\right),
\end{equation}
which is subsequently smoothed using a box filter to suppress residual high-frequency Voronoi artifacts.

\subsubsection{Direction Field}
The direction field encodes a spatially coherent orientation map over the region interior, guiding procedural stroke alignment, cross-hatching, and brush texture synthesis. For each boundary pixel $\mathbf{e}$, we estimate the local tangent direction using the PCA-smoothed tangent vector $\mathbf{t}=(t_x,t_y)$. The corresponding orientation angle is computed as $\phi = \operatorname{atan2}(t_y, t_x) \in (-\pi, \pi]$. This continuous angle is quantized into an 8-bit integer channel via:
\begin{equation}
    M_{\text{dir}}(\mathbf{e}) = \left\lfloor \frac{\phi + \pi}{2\pi} \times 255 \right\rceil,
\end{equation}
where $\lfloor \cdot \rceil$ denotes the nearest-integer rounding function. The quantized boundary orientations are propagated throughout the region interior using the same exponential distance falloff formulation.

\subsection{Procedural Stylization}
We implement a procedural stylization pipeline in Open Shading Language (OSL). The pipeline synthesizes the final texture and artistic appearance of each segmented region using the dense geometric descriptor fields introduced in Section~\ref{sec:geometry}. Guided by these descriptors, the pipeline generates four complementary non-photorealistic rendering primitives: contour-aligned strokes, stochastic dot stippling, high-frequency Gabor noise~\cite{GLLD12}, and quantized dithered shading. For each segmented region, the shader takes the semantically predicted HSL base color as input and dynamically modulates its components using the corresponding geometric descriptor fields. This decoupled design separates semantic color assignment from procedural texture synthesis. It enables perceptually plausible, stylistically diverse renderings that capture the natural variations of hand-drawn artwork.

\subsubsection{Direction-Field Sampling}
The direction field, $M_{\text{dir}}$, is provided to the shader as a normalized scalar texture map. Each value $a(\mathbf{p}) \in [0,1]$ encodes the local boundary orientation at the shader evaluation point $\mathbf{p}=(x,y)^\top$. The scalar value is decoded into a continuous orientation angle $\theta(\mathbf{p})$ and subsequently mapped to the corresponding unit tangent vector $\mathbf{u}(\mathbf{p})$ via:
\begin{equation}
    \theta(\mathbf{p}) = 2\pi a(\mathbf{p}) - \pi, \quad \mathbf{u}(\mathbf{p}) = \begin{pmatrix} \cos\theta(\mathbf{p}) \\ \sin\theta(\mathbf{p}) \end{pmatrix}.
\end{equation}
The resulting tangent vector provides directional guidance throughout the subsequent shader stages. Specifically, it biases stroke orientations and anisotropic sampling by shifting the local sampling position according to $\mathbf{p}'=\mathbf{p}-\frac{1}{2}\mathbf{u}(\mathbf{p})$, thereby ensuring that the synthesized textures remain aligned with the underlying contour geometry.

\subsubsection{Direction-Guided Warping}
Texture coordinates are spatially warped according to the local direction field, producing coherent texture patterns that follow the underlying contour geometry. The resulting directional warp operator applied to a texture channel $I$ is defined as:
\begin{equation}
    \mathrm{warp}(I, s, \mathbf{u}, \alpha) = I\big(\mathbf{p} + \mathbf{u}(\mathbf{p}) \cdot s \cdot \alpha\big),
\end{equation}
where $\alpha$ represents a global warp magnitude coefficient, and $s$ denotes either a uniform scaling scalar or a secondary spatially varying noise lookup. This transformation warps procedural textures to follow the underlying contour geometry instead of the standard image-space axes.

\subsubsection{Stochastic Primitives}
To synthesize coarse sketch strokes, we stochastically distribute drawing primitives over a regular lattice grid. Each grid cell, indexed by the lattice coordinate $(i,j)$, independently generates a drawing primitive if the localized hash function $h$ satisfies $h(\mathbf{x}{ij}) < \tau_p$. Here, $\mathbf{x}{ij}$ denotes the grid cell coordinate, $h(\cdot)$ is a deterministic hash function, and $\tau_p=0.8$ controls the primitive spawning density. The spatial center $\mathbf{q}$ and radius $r$ of each generated primitive are subsequently perturbed using Simplex noise, $\mathcal{N}$, according to:
\begin{equation}
    \mathbf{q} = \mathbf{x}_{ij} + \eta\mathcal{N}(\mathbf{x}_{ij}, \xi), \quad r = r_0\mathcal{N}\left(\frac{\mathbf{x}_{ij}}{s}\right),
\end{equation}
where $\eta$ controls the magnitude of the spatial displacement, $\xi$ acts as an absolute random seed, $r_0$ is the base primitive radius, and $s$ is a spatial noise scale parameter.

Contour-aligned strokes are generated using an identical structural logic over a coarser grid arrangement. The line stroke endpoints are mapped as:
\begin{equation}
    \mathbf{x}_0 = (\Delta_x \cdot i, \; \Delta_y \cdot j)^\top, \quad \mathbf{x}_1 = \mathbf{x}_0 + \ell\mathbf{u}(\mathbf{x}_0)\mathcal{N}(\mathbf{x}_0, \xi),
\end{equation}
where $\Delta_x$ and $\Delta_y$ represent cell step dimensions, $\ell$ controls the target stroke length.

\subsubsection{Quantization and Remapping}
To mimic traditional hand-inked artwork and prevent perfectly smooth digital gradients, the lightness channel is discretely quantized into a fixed number of tonal levels. Given a targeted number of tonal shading levels $L \in \mathbb{Z}^+$, the quantization interval step is defined as $\delta = 1 / (L - 1)$, and the quantized scalar value $Q(v, L)$ is expressed as:
\begin{equation}
    Q(v, L) = \left\lfloor \frac{v}{\delta} \right\rfloor \cdot \delta.
\end{equation}
To ensure consistent data propagation across shader stages, quantities are transformed between coordinate bounds using an affine rescaling operator:
\begin{equation}
    \mathrm{remap}(v, A, B, m, M) = m + (M - m)\frac{v - A}{B - A},
\end{equation}
which linearly remaps the input scalar $v$ from an original domain $[A, B]$ to a targeted shader value range $[m, M]$.

\subsubsection{Structured Variance via Gabor Noise}
To generate high-frequency surface details, such as paper grain and charcoal tooth, we construct a multi-frequency Gabor noise field by summing spectral components evaluated at different frequencies:
\begin{equation}
    \mathcal{G}(\mathbf{p}) = \sum_{k} w_k G\left(\frac{\mathbf{p}}{\lambda_k}, \theta_k\right),
\end{equation}
where $G(\cdot)$ represents a structural Gabor noise kernel evaluated at a local spatial frequency scale $\lambda_k$ and localized orientation $\theta_k$, modulated by blending weights $w_k$.

\subsubsection{HSL Composition}
The individual shader components are composited directly in the HSL color space to produce the final stylized appearance of each segmented region. Specifically, the predicted base lightness, $\mathcal{L}0$, is multiplicatively modulated by the multi-scale Gabor noise field, $\mathcal{G}$, the propagated shadow field, $M{\text{shadow}}$, and the boundary curvature field, $M_{\text{curvature}}$. Prior to composition, these descriptor fields are remapped to the desired intensity range:
\begin{equation}
    \mathcal{L}(\mathbf{p}) = \mathcal{L}_0 \cdot \mathcal{G}(\mathbf{p}) \cdot M_{\text{shadow}}(\mathbf{p}) \cdot M_{\text{curvature}}(\mathbf{p}).
\end{equation}
The stochastic dot and stroke primitive layers are then composited onto the lightness channel to enhance structural detail and artistic variation. To emulate the physical diffusion of ink and pigment, the saturation channel, $\mathcal{S}$, is locally modulated according to the primitive density, introducing subtle chromatic variations.

\begin{algorithm}[ht]
\caption{TexSketch: Procedural Sketch Colorization Approach}
\label{alg:algo}
    \begin{algorithmic}[1]
        \State $\mathcal{S} \gets \text{Segment}(I)$
        \State $\mathcal{S} \gets \{s \in \mathcal{S} \mid s \neq \text{background}\}$
\ForAll{$s \in \mathcal{S}$}
    \State $c_s \gets \text{VLMContext}(s)$
    \State $M_{\text{dist}}, M_{\text{curv}}, M_\text{shadow}, M_\text{dir} \gets \text{GenerateDescriptors}(\partial s)$
    \State $P \gets \text{CreatePrimitives}(M_\text{dir})$
    \State $M_{\text{dist}}, M_{\text{curv}}, M_\text{shadow} \gets \text{Quantize}(M_{\text{dist}}, M_{\text{curv}}, M_\text{shadow})$
    \State $s_c \gets \text{Color}(s,c_s,P,M_{\text{dist}}, M_{\text{curv}}, M_\text{shadow}, M_\text{dir})$
    \State $s_c \gets s_c \odot s$
\EndFor
\State $I_c \gets \text{ScreenBlend}(\{s_c : s\in\mathcal{S}\})$
\State $I_c \gets \text{ReplaceBlack}(I_c)$
\State $I_c \gets I_c - \text{MergedBoundary}(\mathcal{S})$
\State $I_c \gets \text{Composite}( I_c,  \text{ForegroundMask}(\mathcal{S}) \odot I$)
\State \Return $I_c$

    \end{algorithmic}
\end{algorithm}

\begin{figure*}[htbp]
  \centering
    \setlength{\tabcolsep}{0pt}
    \small
    \begin{tabular}{ccccccccc}
        & \textbf{Input sketch} & \textbf{Color sketch} & \textbf{Input sketch} & \textbf{Color sketch} & \textbf{Input sketch} & \textbf{Color sketch}\\
        & \raisebox{-0.5\height}{\includegraphics[width=.15\linewidth]{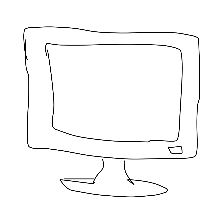}}
        & \raisebox{-0.5\height}{\includegraphics[width=.15\linewidth]{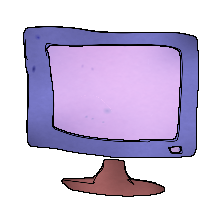}}
        & \raisebox{-0.5\height}{\includegraphics[width=.15\linewidth]{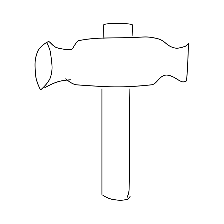}}
        & \raisebox{-0.5\height}{\includegraphics[width=.15\linewidth]{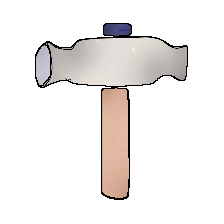}}
        & \raisebox{-0.5\height}{\includegraphics[width=.15\linewidth]{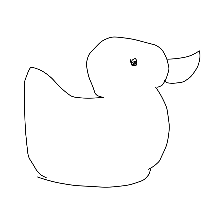}}
        & \raisebox{-0.5\height}{\includegraphics[width=.15\linewidth]{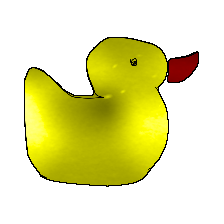}} \\

        & \raisebox{-0.5\height}{\includegraphics[width=.15\linewidth]{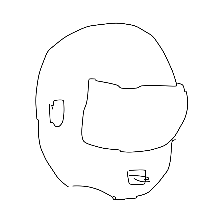}}
        & \raisebox{-0.5\height}{\includegraphics[width=.15\linewidth]{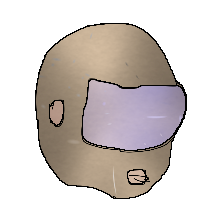}}
        & \raisebox{-0.5\height}{\includegraphics[width=.15\linewidth]{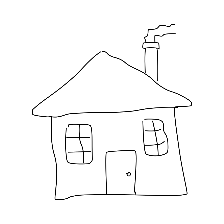}}
        & \raisebox{-0.5\height}{\includegraphics[width=.15\linewidth]{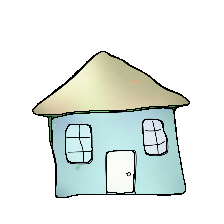}}
        & \raisebox{-0.5\height}{\includegraphics[width=.15\linewidth]{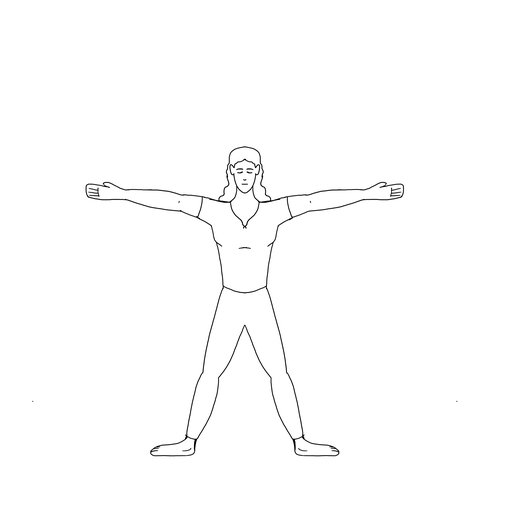}}
        & \raisebox{-0.5\height}{\includegraphics[width=.15\linewidth]{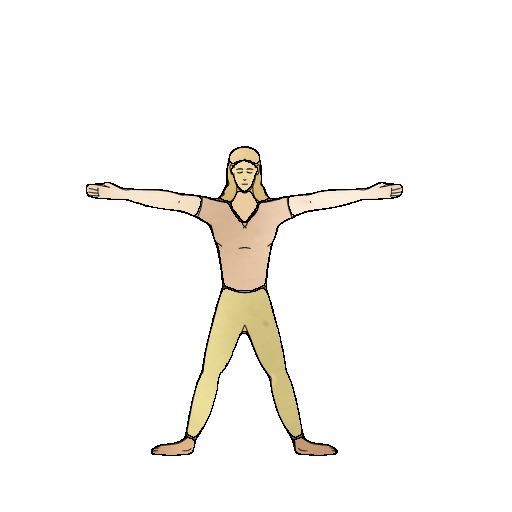}} \\

        & \raisebox{-0.5\height}{\includegraphics[width=.15\linewidth]{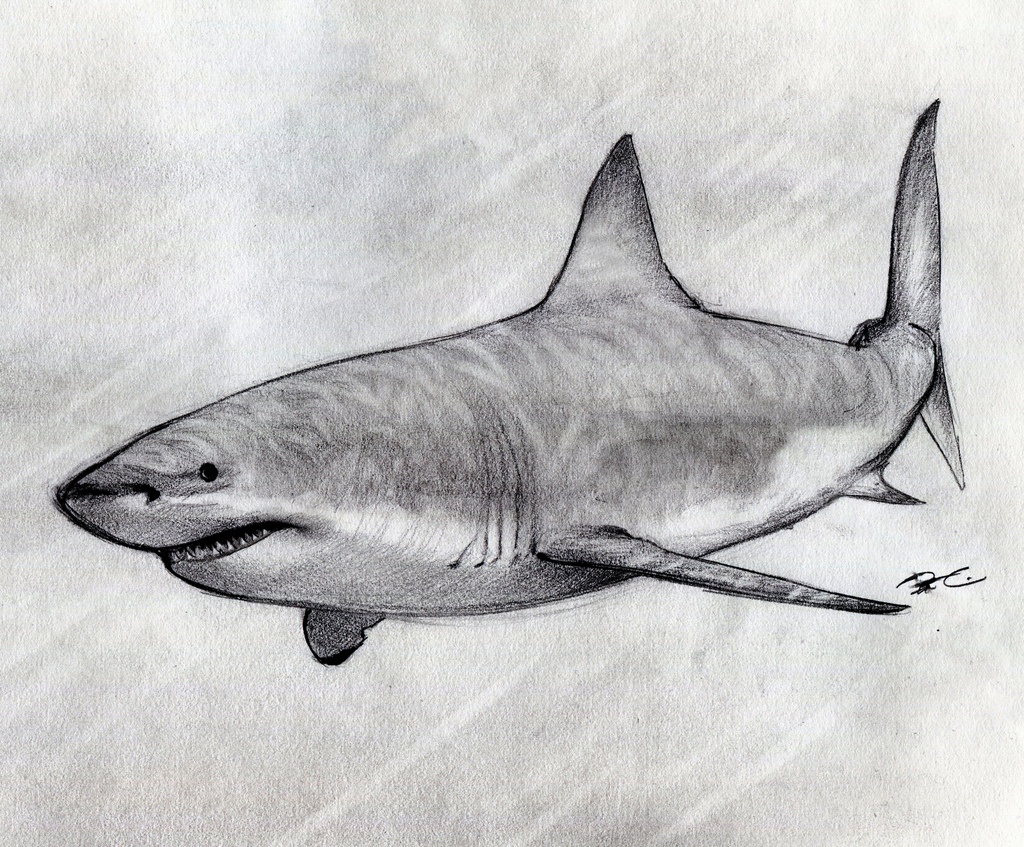}}
        & \raisebox{-0.5\height}{\includegraphics[width=.15\linewidth]{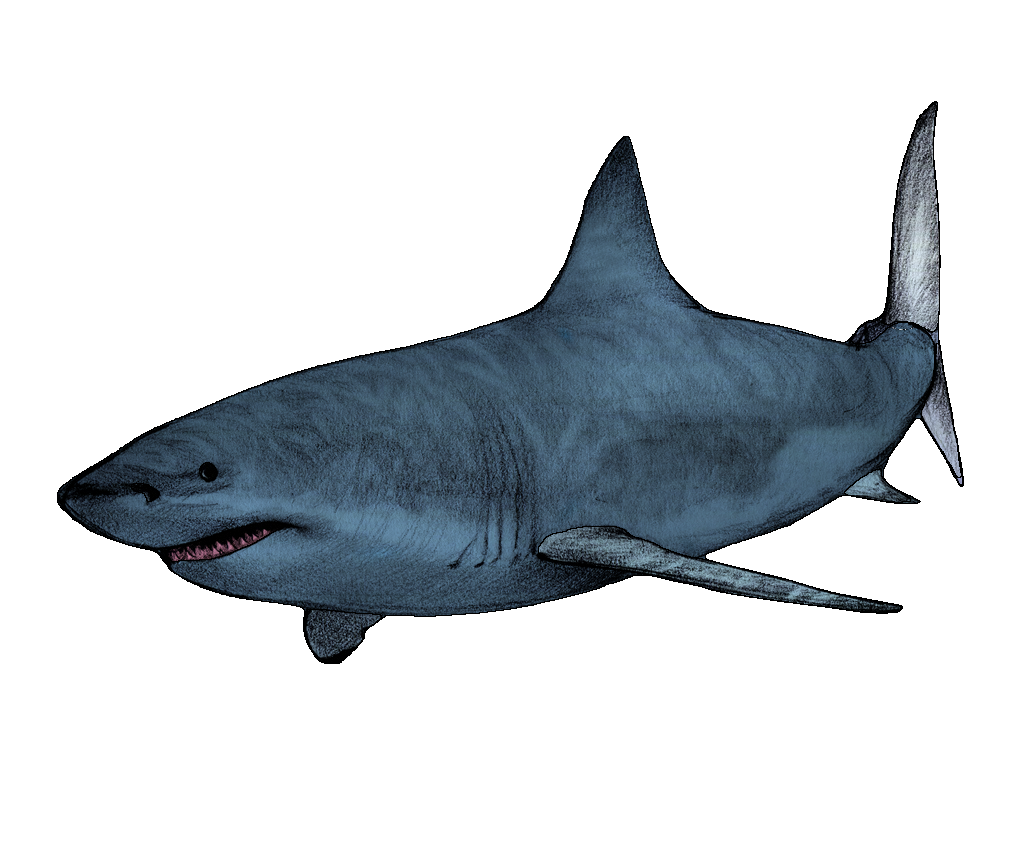}}
        & \raisebox{-0.5\height}{\includegraphics[width=.15\linewidth]{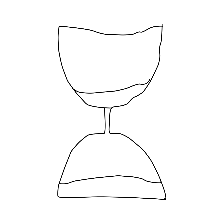}}
        & \raisebox{-0.5\height}{\includegraphics[width=.15\linewidth]{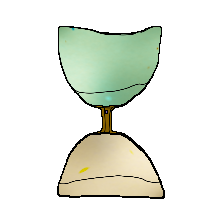}}
        & \raisebox{-0.5\height}{\includegraphics[width=.15\linewidth]{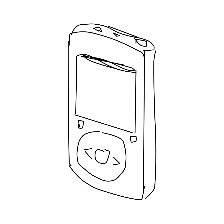}}
        & \raisebox{-0.5\height}{\includegraphics[width=.15\linewidth]{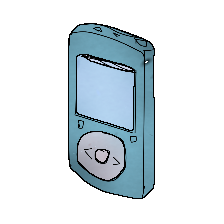}} \\

        & \raisebox{-0.5\height}{\includegraphics[width=.15\linewidth]{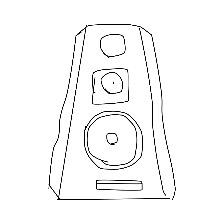}}
        & \raisebox{-0.5\height}{\includegraphics[width=.15\linewidth]{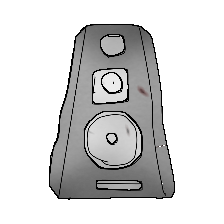}} 
        & \raisebox{-0.5\height}{\includegraphics[width=.15\linewidth]{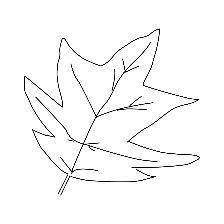}}
        & \raisebox{-0.5\height}{\includegraphics[width=.15\linewidth]{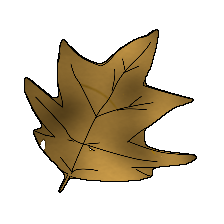}}
        & \raisebox{-0.5\height}{\includegraphics[width=.15\linewidth]{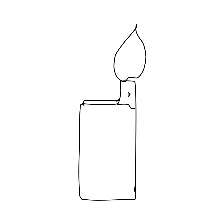}}
        & \raisebox{-0.5\height}{\includegraphics[width=.15\linewidth]{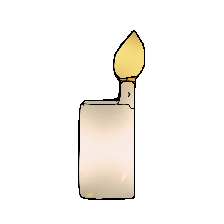}}
    \end{tabular}
    \caption{Visual results of the proposed TexSketch framework on sketches from ImageNet~\cite{wang2019learning} and TU-Berlin~\cite{eitz2012humans} datasets.}
    \label{fig:our_results}
\end{figure*}

\subsection{Compositing and Post-Processing}
Following the independent stylization of each segmented region, the resulting layers are composited into a unified rendered image. Each layer is first multiplied by its corresponding segmentation mask to ensure clean, anti-aliased boundary transitions. The masked assets are combined over a clean, white paper canvas background using standard alpha-blended over-compositing. The original sketch contour mask is subsequently overlaid onto the composite image, reinforcing the input line art by sharpening region boundaries and correcting minor color bleeding introduced during procedural texture synthesis. As a final refinement, the extracted segment masks are used to blend the original sketch with the stylized rendering, preserving fine line details and maintaining the structural fidelity of the input sketch.

\begin{figure}[htbp]
  \centering
    \setlength{\tabcolsep}{4pt}
    \scalebox{0.8}{
    \small
    \begin{tabular}{ccccccc}
        \vspace{0.1cm}
        & \textbf{Input sketch} & \textbf{SketchDeco~\cite{utintu2026sketchdeco}} & \textbf{ColorizeDiffusionXL~\cite{yan2026towards}} & \textbf{Ours} \\ \vspace{0.1cm}
        & \raisebox{-0.2\height}{\includegraphics[width=.2\linewidth]{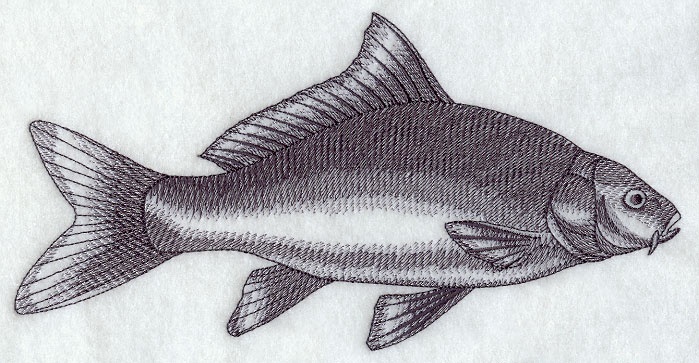}}
        & \raisebox{-0.2\height}{\includegraphics[width=.2\linewidth]{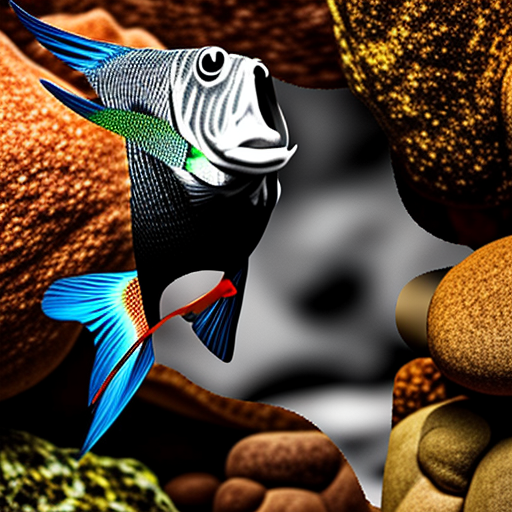}}
        & \raisebox{-0.2\height}{\includegraphics[width=.2\linewidth]{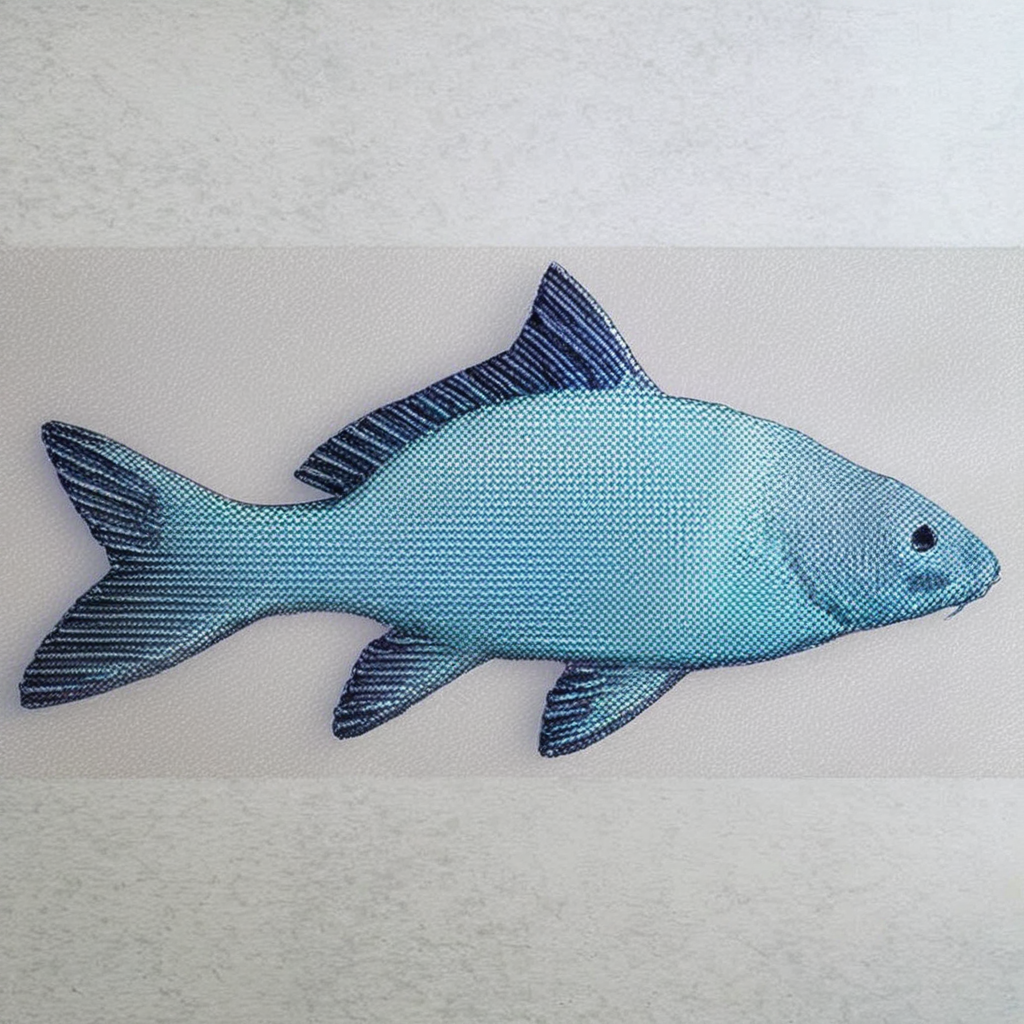}}
        & \raisebox{-0.2\height}{\includegraphics[width=.2\linewidth]{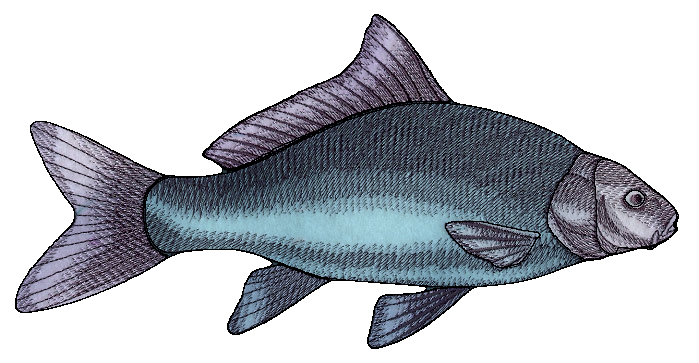}} \\
        \vspace{0.1cm}
        
        & \raisebox{-0.3\height}{\includegraphics[width=.2\linewidth]{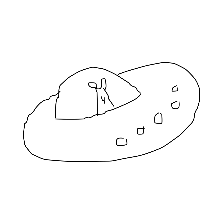}}
        & \raisebox{-0.3\height}{\includegraphics[width=.2\linewidth]{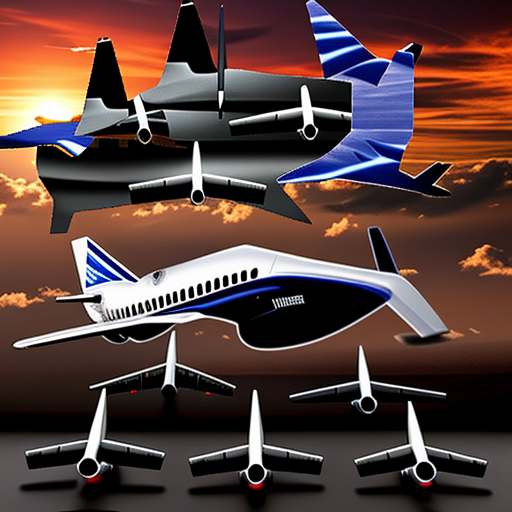}}
        & \raisebox{-0.3\height}{\includegraphics[width=.2\linewidth]{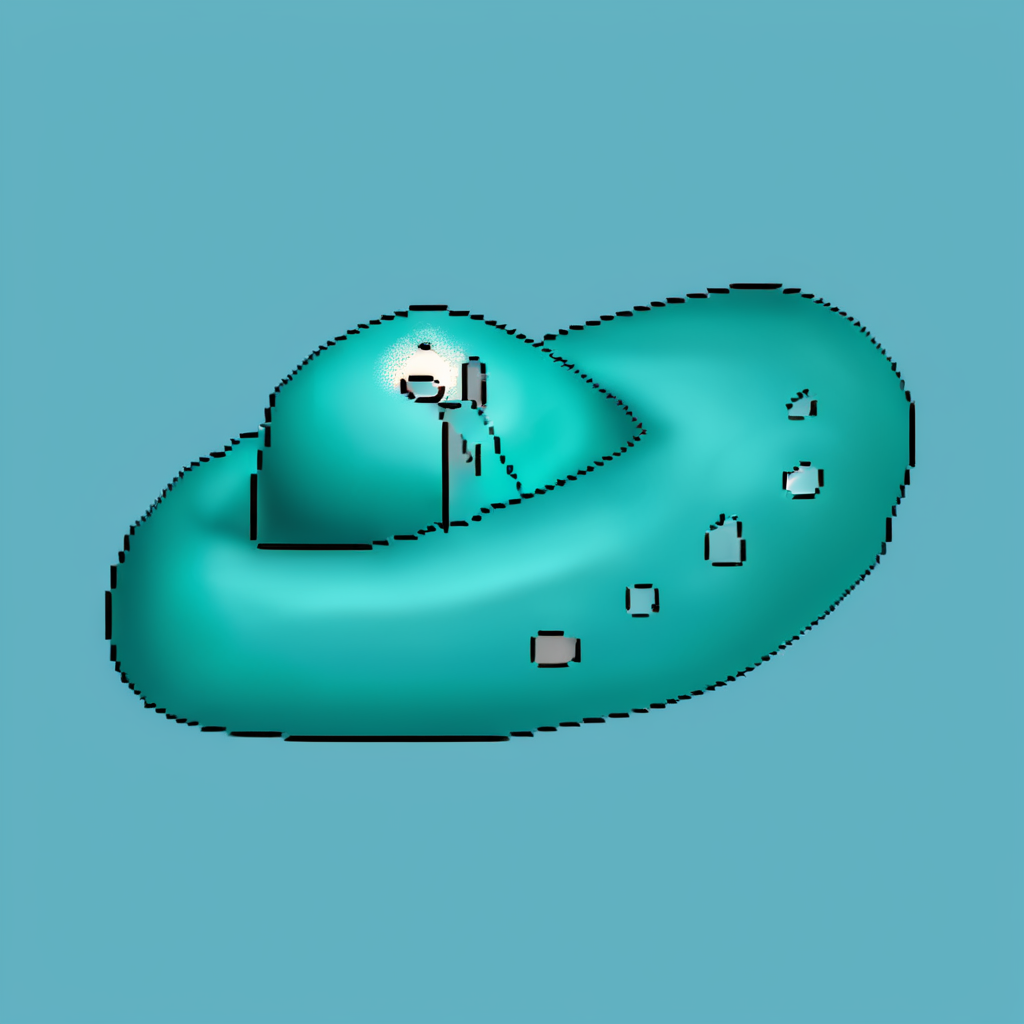}}
        & \raisebox{-0.3\height}{\includegraphics[width=.2\linewidth]{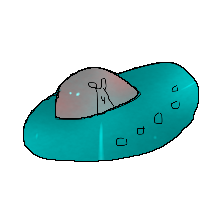}} \\
        \vspace{0.1cm}

        & \raisebox{-0.2\height}{\includegraphics[width=.2\linewidth]{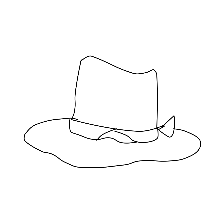}}
        & \raisebox{-0.2\height}{\includegraphics[width=.2\linewidth]{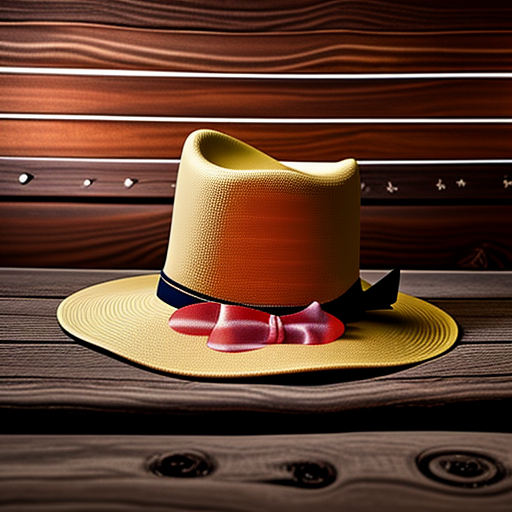}}
        & \raisebox{-0.2\height}{\includegraphics[width=.2\linewidth]{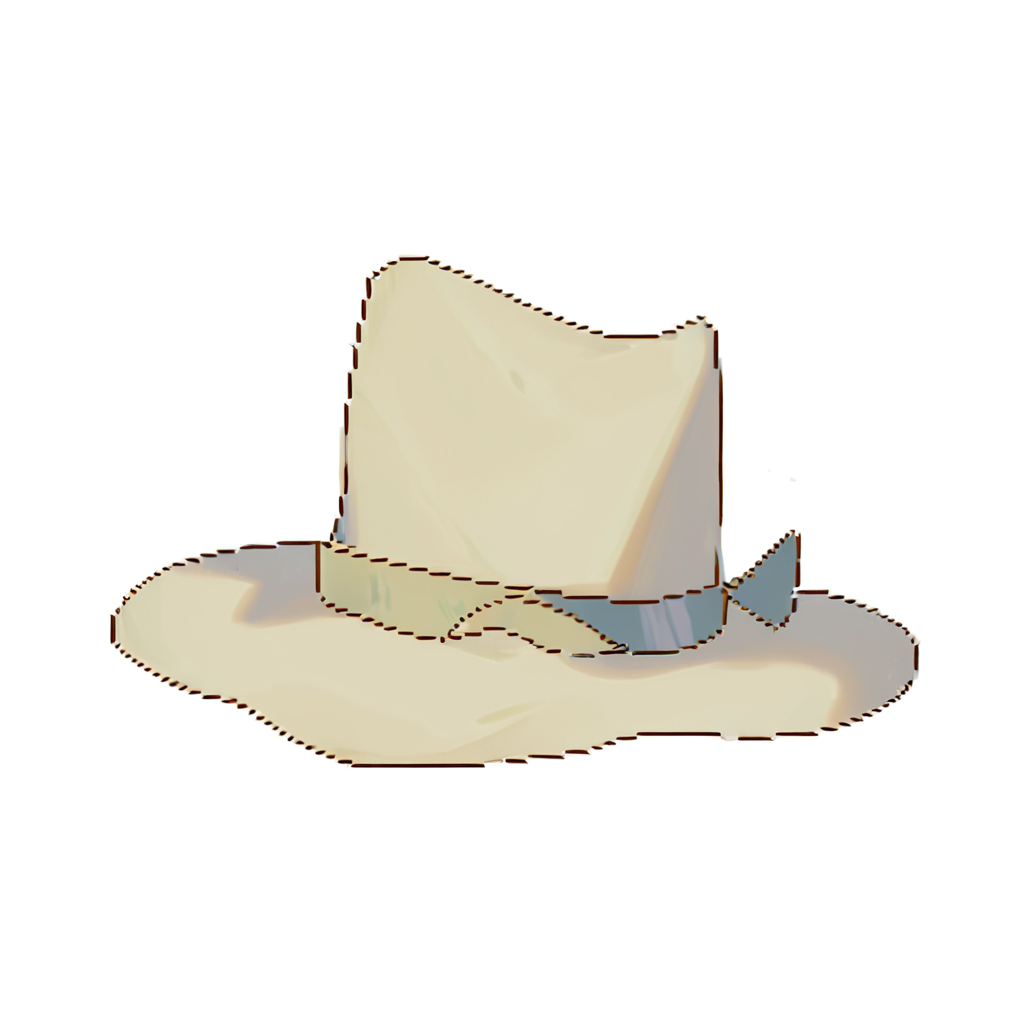}}
        & \raisebox{-0.2\height}{\includegraphics[width=.2\linewidth]{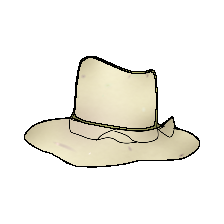}} \\
        \vspace{0.1cm}

        & \raisebox{-0.2\height}{\includegraphics[width=.2\linewidth]{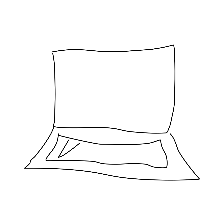}}
        & \raisebox{-0.2\height}{\includegraphics[width=.2\linewidth]{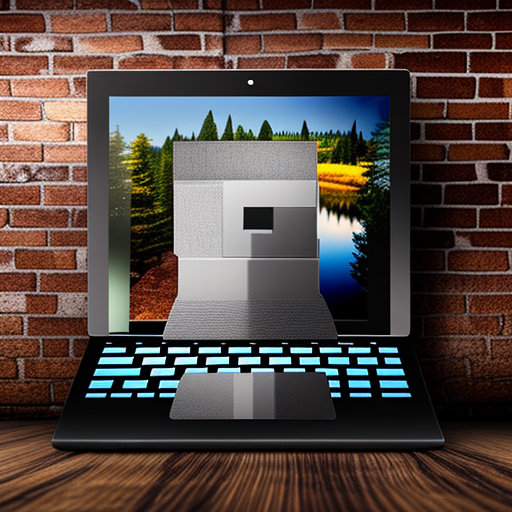}}
        & \raisebox{-0.2\height}{\includegraphics[width=.2\linewidth]{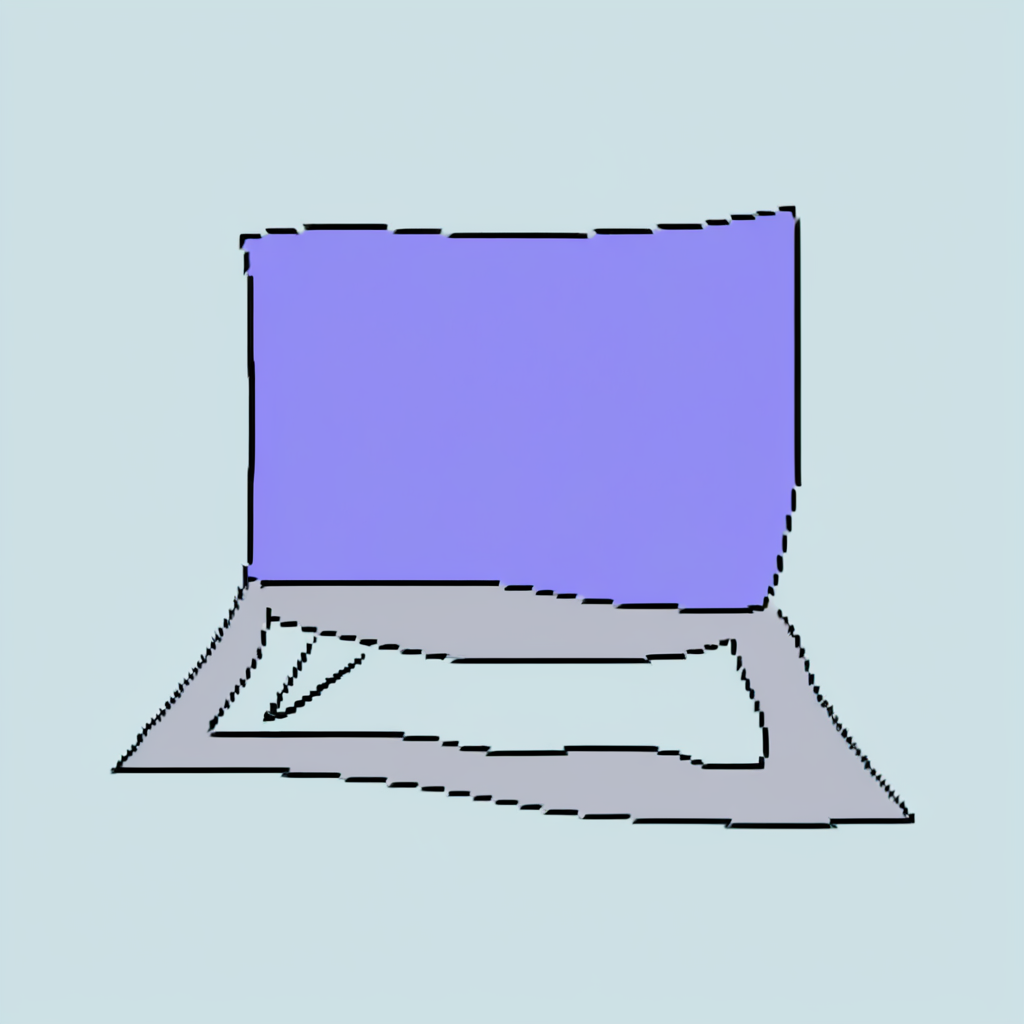}}
        & \raisebox{-0.2\height}{\includegraphics[width=.2\linewidth]{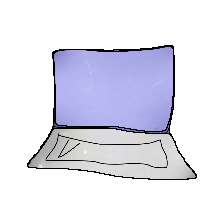}} \\
        \vspace{0.1cm}

        & \raisebox{-0.5\height}{\includegraphics[width=.2\linewidth]{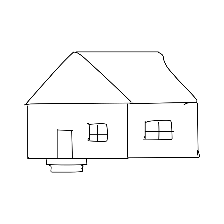}}
        & \raisebox{-0.5\height}{\includegraphics[width=.2\linewidth]{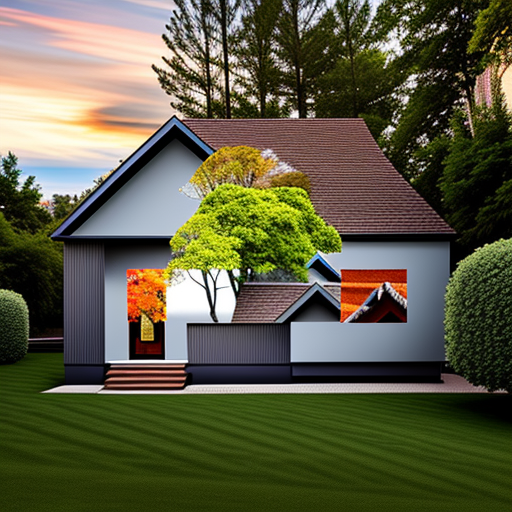}}
        & \raisebox{-0.5\height}{\includegraphics[width=.2\linewidth]{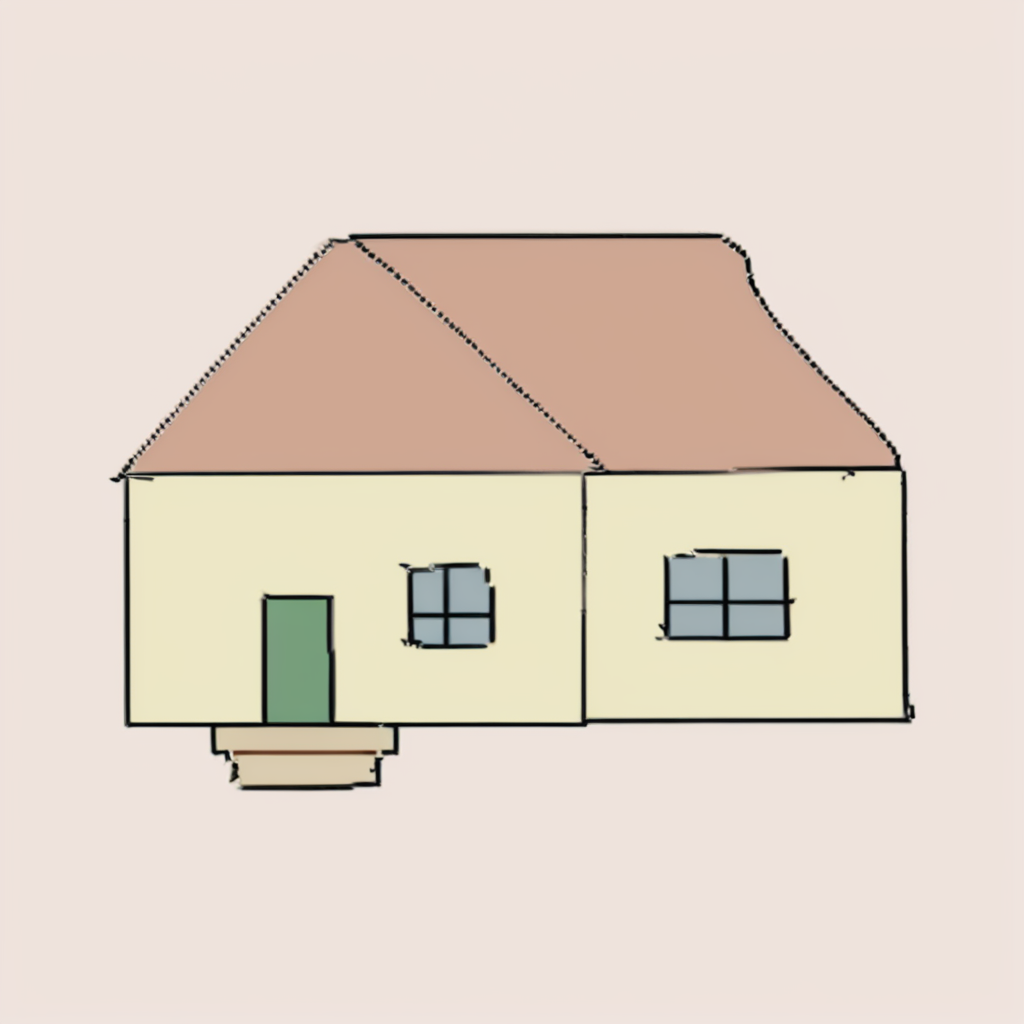}}
        & \raisebox{-0.5\height}{\includegraphics[width=.2\linewidth]{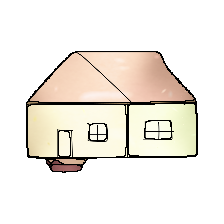}} \\
        \vspace{0.1cm}

        & \raisebox{-0.5\height}{\includegraphics[width=.2\linewidth]{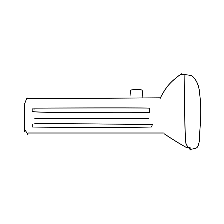}}
        & \raisebox{-0.5\height}{\includegraphics[width=.2\linewidth]{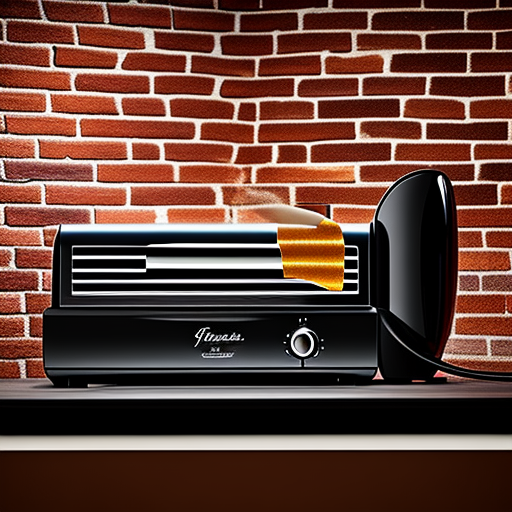}}
        & \raisebox{-0.5\height}{\includegraphics[width=.2\linewidth]{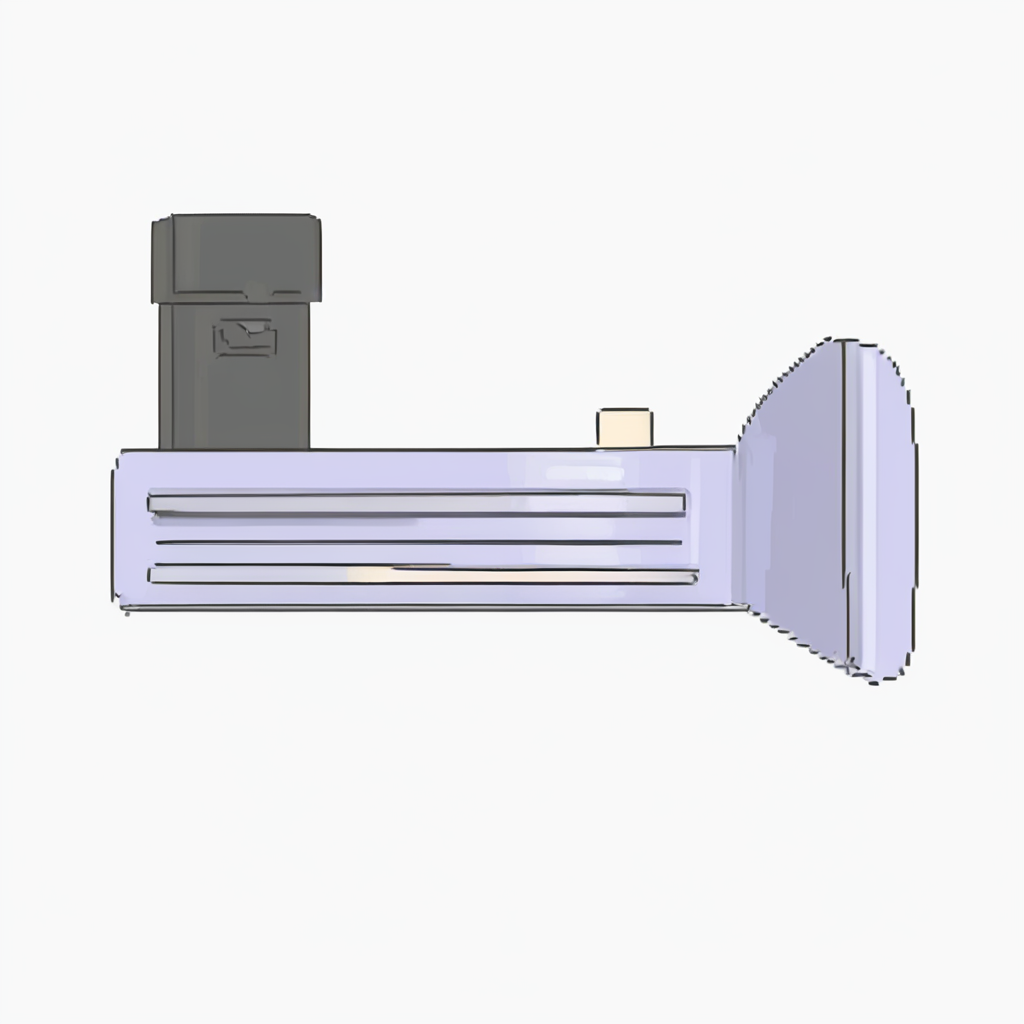}}
        & \raisebox{-0.5\height}{\includegraphics[width=.2\linewidth]{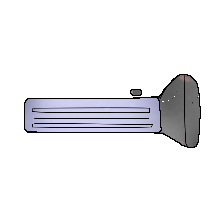}} \\
        \vspace{0.1cm}

        & \raisebox{-0.5\height}{\includegraphics[width=.2\linewidth]{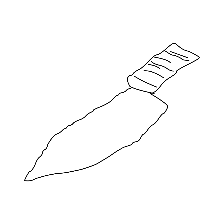}}
        & \raisebox{-0.5\height}{\includegraphics[width=.2\linewidth]{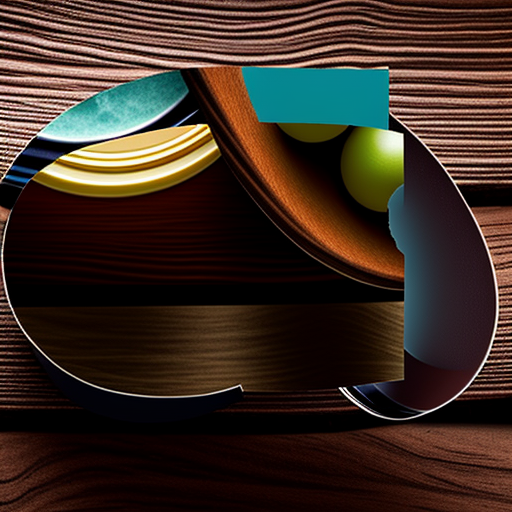}}
        & \raisebox{-0.5\height}{\includegraphics[width=.2\linewidth]{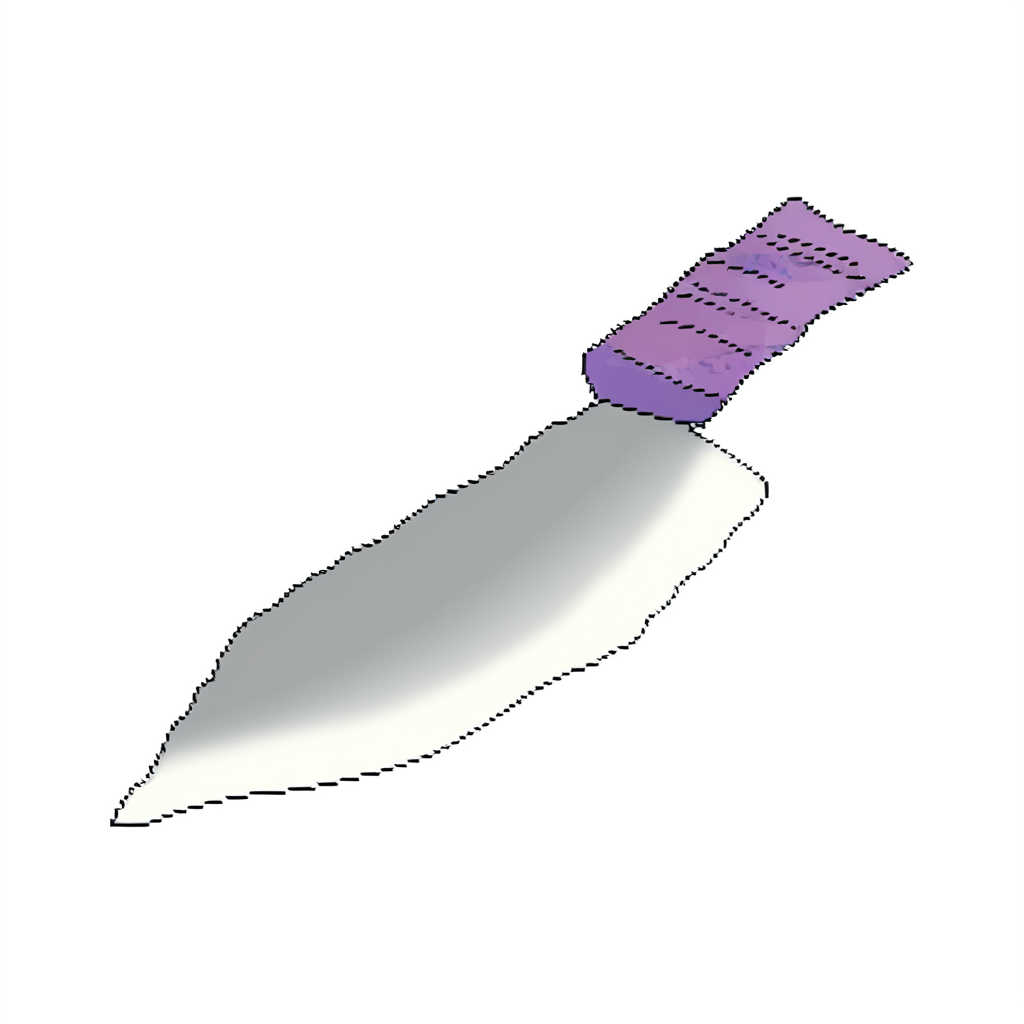}}
        & \raisebox{-0.5\height}{\includegraphics[width=.2\linewidth]{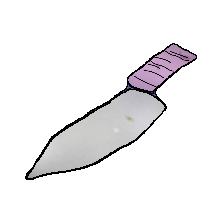}} \\
        \vspace{0.1cm}

        & \raisebox{-0.5\height}{\includegraphics[width=.2\linewidth]{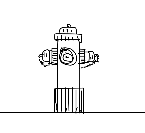}}
        & \raisebox{-0.5\height}{\includegraphics[width=.2\linewidth]{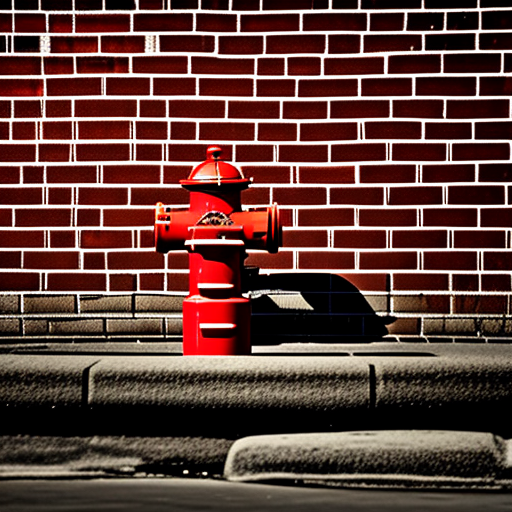}}
        & \raisebox{-0.5\height}{\includegraphics[width=.2\linewidth]{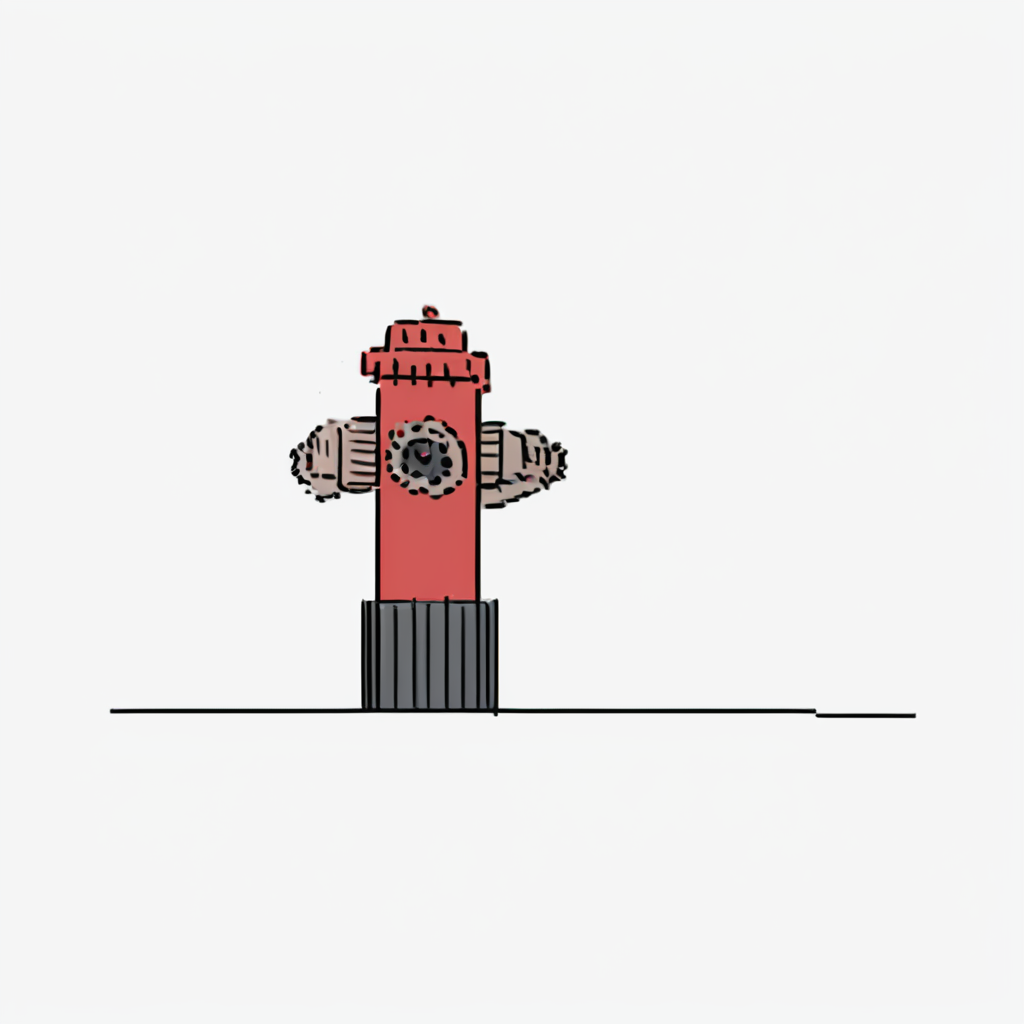}}
        & \raisebox{-0.5\height}{\includegraphics[width=.2\linewidth]
        {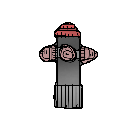}} \\
        \vspace{0.1cm}

        & \raisebox{-0.5\height}{\includegraphics[width=.2\linewidth]{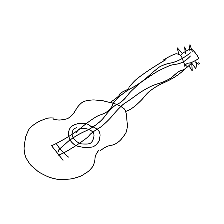}}
        & \raisebox{-0.5\height}{\includegraphics[width=.2\linewidth]{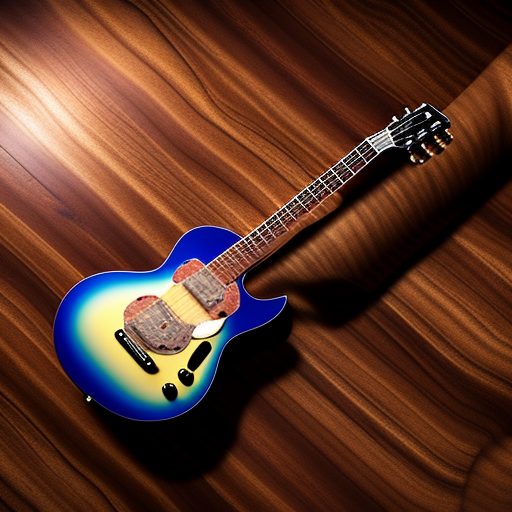}}
        & \raisebox{-0.5\height}{\includegraphics[width=.2\linewidth]{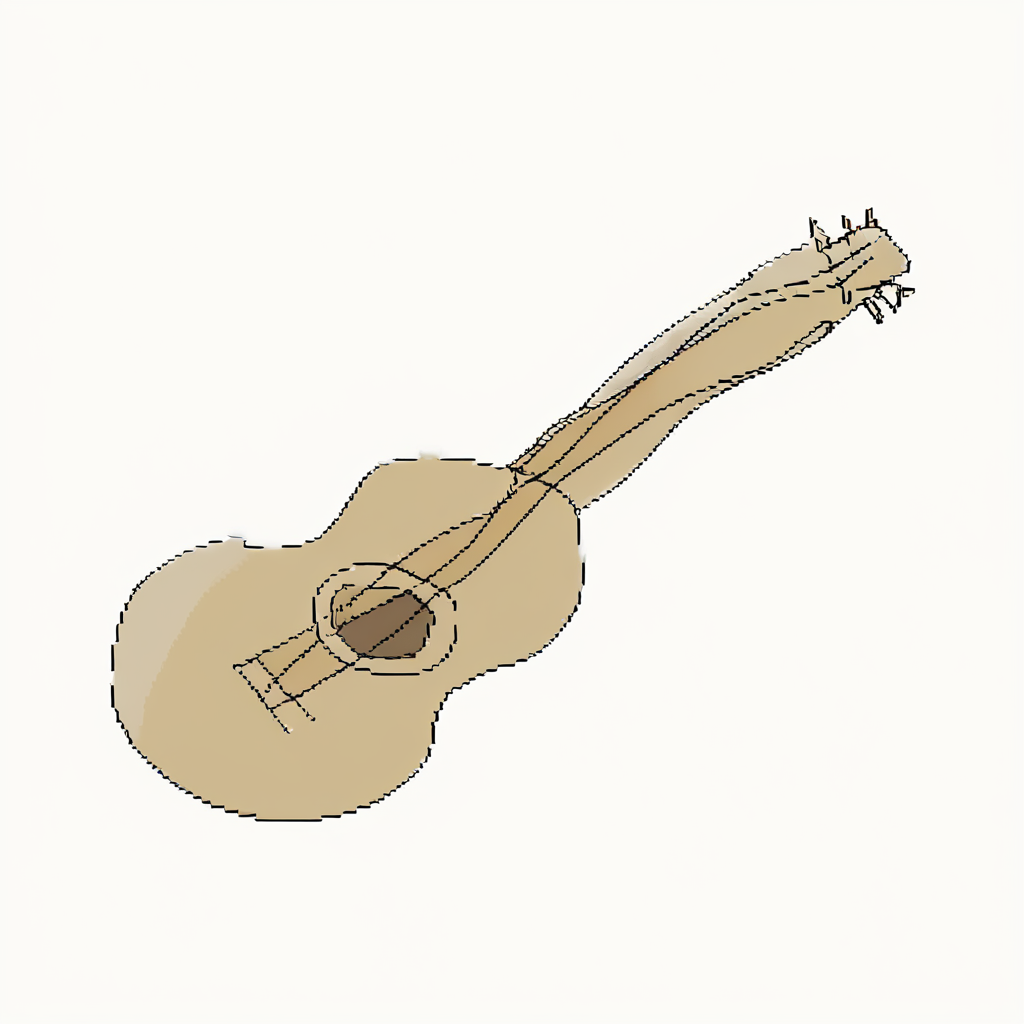}}
        & \raisebox{-0.5\height}{\includegraphics[width=.2\linewidth]{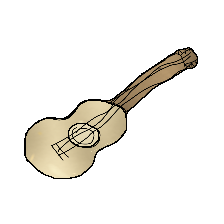}} \\
        \vspace{0.1cm}

        & \raisebox{-0.5\height}{\includegraphics[width=.2\linewidth]{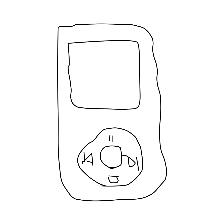}}
        & \raisebox{-0.5\height}{\includegraphics[width=.2\linewidth]{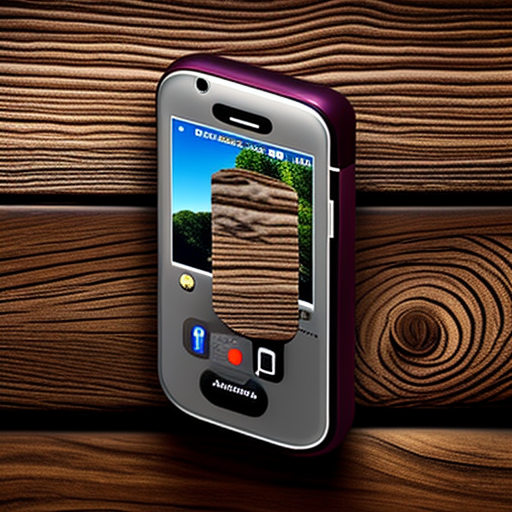}}
        & \raisebox{-0.5\height}{\includegraphics[width=.2\linewidth]{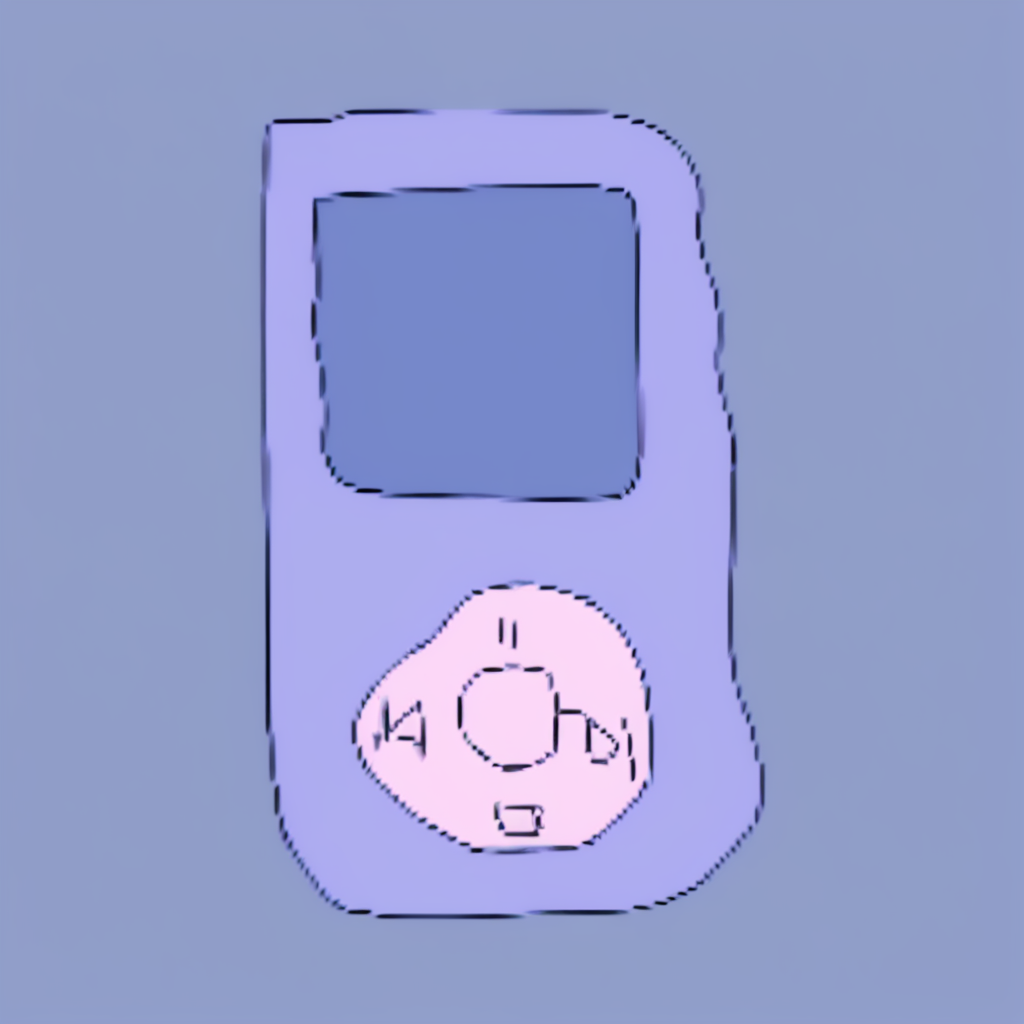}}
        & \raisebox{-0.5\height}{\includegraphics[width=.2\linewidth]{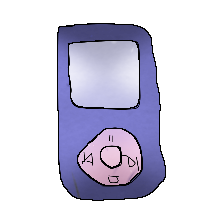}} \\
    \end{tabular}
    }
    \caption{Qualitative comparison of our method with ColorizeDiffusionXL~\cite{yan2026towards} and SketchDeco~\cite{utintu2026sketchdeco}.}
    \label{fig:comparison}
\end{figure}

\section{Results and Comparison}
We evaluate TexSketch on two widely used sketch datasets: ImageNetSketch~\cite{wang2019learning} and TU-Berlin~\cite{eitz2012humans}. 
ImageNetSketch~\cite{wang2019learning} contains structurally detailed sketches derived from natural images, while TU-Berlin~\cite{eitz2012humans} comprises approximately 20,000 freehand sketches across 250 object categories with varying levels of abstraction. Together, these datasets enable the evaluation of the proposed framework on both detailed and highly abstract sketch styles. Algorithm~\ref{alg:algo} outlines the sequence of operations of our proposed framework for generating the final colored sketch. Fig.~\ref{fig:our_results} demonstrates the sketch colorization results produced by the proposed method on a diverse set of arbitrary input sketches. Implementation details are provided in the supplementary material.

We compare the proposed method with state-of-the-art sketch colorization approaches, ColorizeDiffusionXL~\cite{yan2026towards} and SketchDeco~\cite{utintu2026sketchdeco}. The primary objective of TexSketch is to generate visually plausible colored sketches with diverse artistic styles. Fig.~\ref{fig:comparison} provides the visual comparison of our method with ColorizeDiffusionXL~\cite{yan2026towards} and SketchDeco~\cite{utintu2026sketchdeco}. We conduct a subjective user study with 40 participants on 15 distinct samples to assess the perceptual quality of the generated colored sketches. For each input sketch, participants were shown the outputs generated by all three methods. They were then asked to independently rate each result on a five-point scale, ranging from very poor to excellent (1: "poor"; 2: "fair"; 3: "average"; 4: "good"; 5: "excellent") for every evaluation criterion. 
The evaluation considers five criteria: (i) color coherence and consistency, assessing whether semantically related regions receive balanced and consistent colors; (ii) overall aesthetic appeal, measuring the visual attractiveness of the final rendering; (iii) semantic correctness, evaluating how convincingly the image objects receive semantically plausible colors; (iv) sketch fidelity, measuring how well the generated result preserves the original sketch structure and fine line details; and (v) artifact level, assessing the presence of undesirable visual artifacts such as color bleeding, texture discontinuities, or rendering noise.

\begin{table}[t]
    \centering
    \scriptsize
    \caption{Comparison of the proposed method with state-of-the-art sketch colorization approaches. Mean opinion scores from the user study are reported for each evaluation criterion, where higher scores indicate better perceptual quality.}
    \setlength{\tabcolsep}{4pt}
    \label{tab:comparison}
    \begin{tabular}{l||ccccc}
        \toprule
        \multirow{2}{*}{Methods} & \makecell{Color\\consistency} & \makecell{Overall\\aesthetics} &  \makecell{Semantic\\correctness} & \makecell{Sketch\\fidelity} & \makecell{Artifact\\level} \\
        \midrule
        ColorizeDiffusionXL~\cite{yan2026towards} & 3.59 & 3.45 & 3.17 & 3.37 & 3.34 \\
        SketchDeco~\cite{utintu2026sketchdeco} & 3.23 & 3.20 & 2.78 & 2.96 & 2.99\\
        Ours & \textbf{3.66} & \textbf{3.55} & \textbf{3.47} & \textbf{3.70} & \textbf{3.66} \\
        \bottomrule
    \end{tabular}
\end{table}

\begin{figure}[t]
    \centering
    \begin{tikzpicture}
        \begin{axis}[
            ybar=1pt,                     
            bar width=7pt,
            width=\linewidth,
            height=4.5cm,
            ymin=1, ymax=5,               
            ylabel={Mean rating (1--5)},
            symbolic x coords={Color cons., Aesthetics, Semantic corr., Sketch fid., Artifacts},
            xtick=data,
            x tick label style={font=\scriptsize, rotate=20, anchor=east},
            tick label style={font=\scriptsize},
            label style={font=\small},
            ymajorgrids=true,
            grid style={white!85!black, line width=0.5pt},
            axis line style={white!60!black},
            tick style={white!60!black},
            legend style={
                font=\scriptsize, 
                at={(0.5, 1.15)}, 
                anchor=north, 
                legend columns=-1,
                draw=none,                 
                fill=none
            },
            enlarge x limits=0.15,
            error bars/y dir=both,
            error bars/y explicit,
            error bars/error mark options={
                rotate=90,
                mark size=2pt,
                line width=0.3pt,
                solid
            },
        ]
        \addplot[fill=sblue, draw=sblue!80!black] coordinates {
            (Color cons., 3.59) += (0, 0.10) -= (0, 0.11)
            (Aesthetics,  3.45) += (0, 0.10) -= (0, 0.10)
            (Semantic corr., 3.17) += (0, 0.11) -= (0, 0.11)
            (Sketch fid., 3.37) += (0, 0.11) -= (0, 0.11)
            (Artifacts,   3.34) += (0, 0.11) -= (0, 0.11)
        };

        \addplot[fill=sorange, draw=sorange!80!black] coordinates {
            (Color cons., 3.23) += (0, 0.12) -= (0, 0.12)
            (Aesthetics,  3.20) += (0, 0.12) -= (0, 0.12)
            (Semantic corr., 2.78) += (0, 0.13) -= (0, 0.12)
            (Sketch fid., 2.96) += (0, 0.13) -= (0, 0.13)
            (Artifacts,   2.99) += (0, 0.13) -= (0, 0.13)
        };

        \addplot[fill=sgreen, draw=sgreen!80!black] coordinates {
            (Color cons., 3.66) += (0, 0.10) -= (0, 0.10)
            (Aesthetics,  3.55) += (0, 0.10) -= (0, 0.11)
            (Semantic corr., 3.47) += (0, 0.11) -= (0, 0.12)
            (Sketch fid., 3.69) += (0, 0.11) -= (0, 0.11)
            (Artifacts,   3.61) += (0, 0.11) -= (0, 0.11)
        };
        \legend{ColorizeDiffusionXL~\cite{yan2026towards}, SketchDeco~\cite{utintu2026sketchdeco}, Ours}
        \end{axis}
    \end{tikzpicture}
    \caption{Mean opinion score per evaluation criterion for each sketch colorization method (with 95\% confidence interval).}
    \label{fig:mean-rating-bars}
\end{figure}
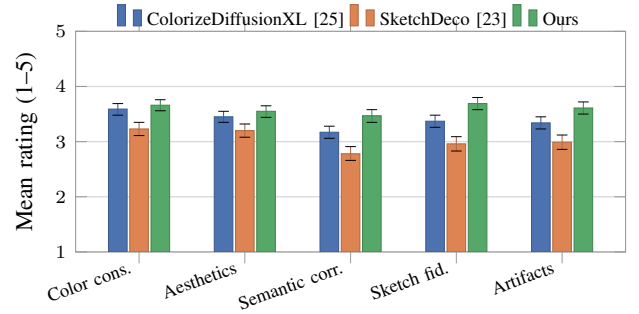

Our proposed method achieves the highest mean opinion score on all five evaluation criteria, demonstrating its ability to generate visually coherent, perceptually convincing and stylistically consistent colored sketches, as shown in Table~\ref{tab:comparison}. In particular, it achieves the largest performance gains in sketch fidelity and artifact level, indicating that the generated results better preserve the structural details of the input sketches while exhibiting fewer visual artifacts, such as color bleeding and texture inconsistencies. Although ColorizeDiffusionXL~\cite{yan2026towards} performs competitively on the color-related criteria, the proposed method consistently exhibit higher scores across all evaluation metrics, including overall aesthetic appeal and semantic correctness, suggesting that its procedurally generated textures more closely resemble traditional hand-drawn artwork. In comparison, SketchDeco~\cite{utintu2026sketchdeco} receives the lowest ratings across all five evaluation criteria, reflecting comparatively weaker performance in terms of visual quality, structural preservation, rendering consistency and preservation.
Fig.~\ref{fig:mean-rating-bars} presents a visual comparison of the mean user ratings for the proposed method and the competing state-of-the-art approaches across all evaluation criteria.

\section{Style Controllability}
\begin{figure}
  \centering
  \setlength{\tabcolsep}{2pt}
  \scalebox{0.9}{
  \small
  \begin{tabular}{cccc}
    \textbf{Input sketch} & \textbf{Base} & \textbf{Cel shading} & \textbf{Watercolor} \\

    \raisebox{-0.5\height}{\includegraphics[width=.25\linewidth]{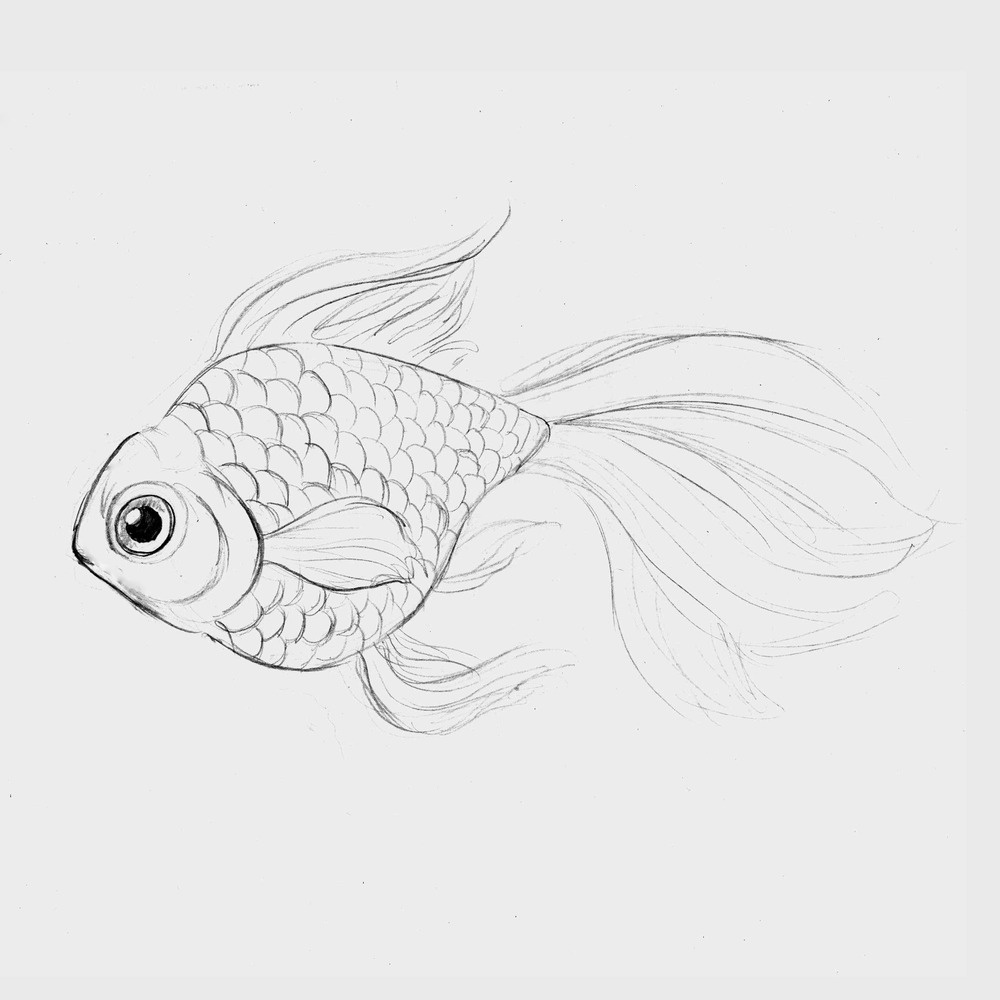}} &
    \raisebox{-0.5\height}{\includegraphics[width=.25\linewidth]{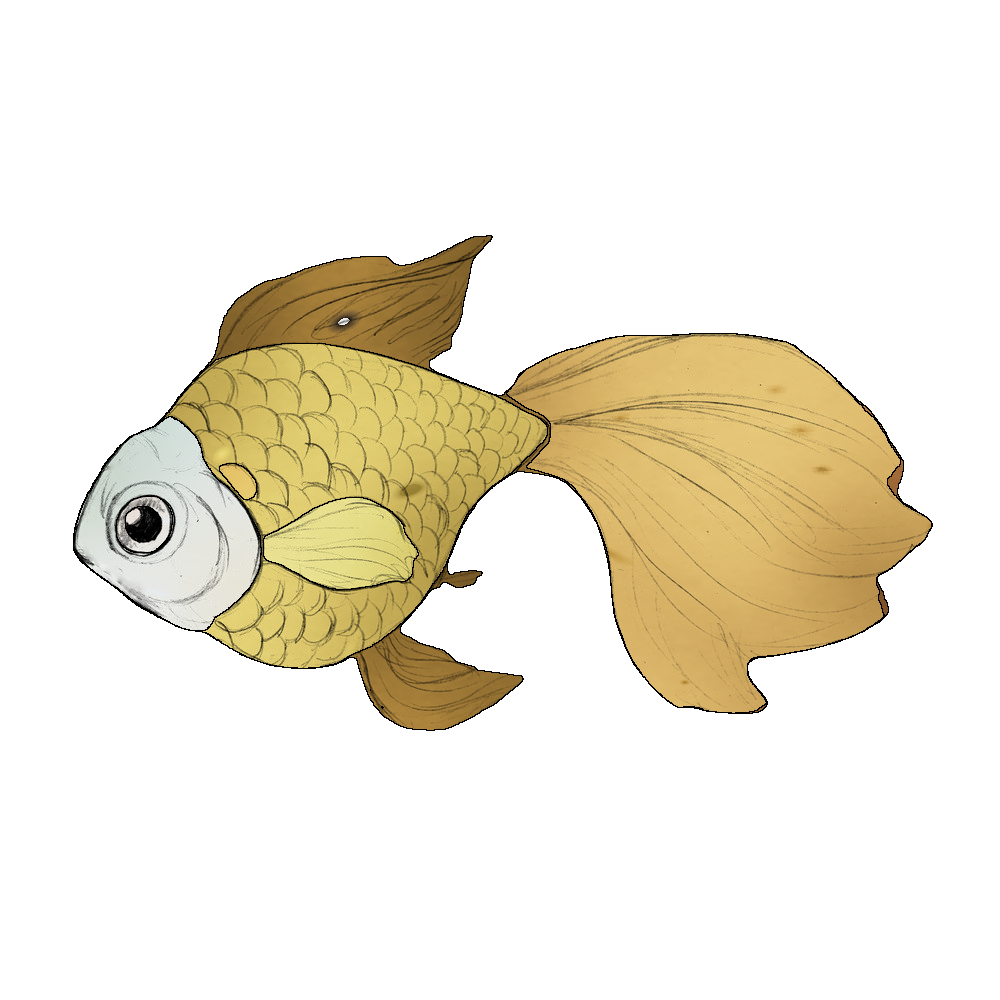}} &
    \raisebox{-0.5\height}{\includegraphics[width=.25\linewidth]{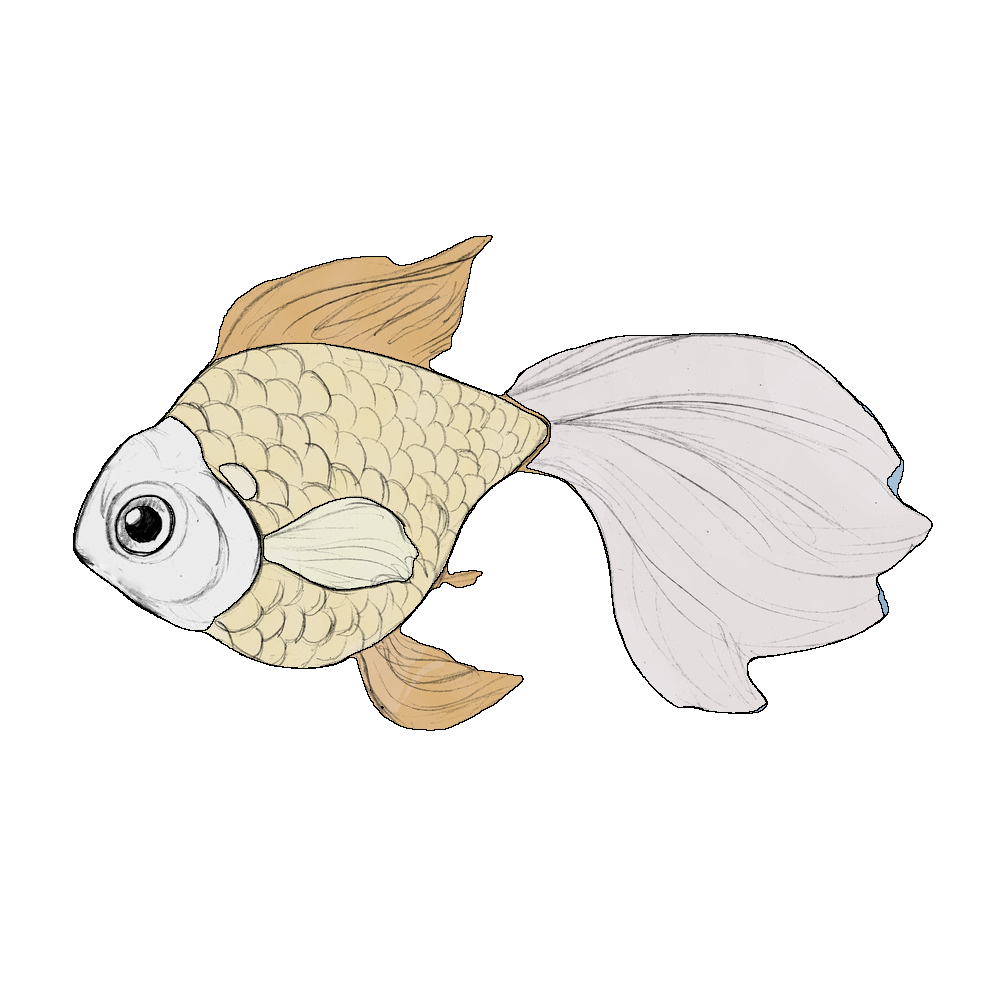}} &
    \raisebox{-0.5\height}{\includegraphics[width=.25\linewidth]{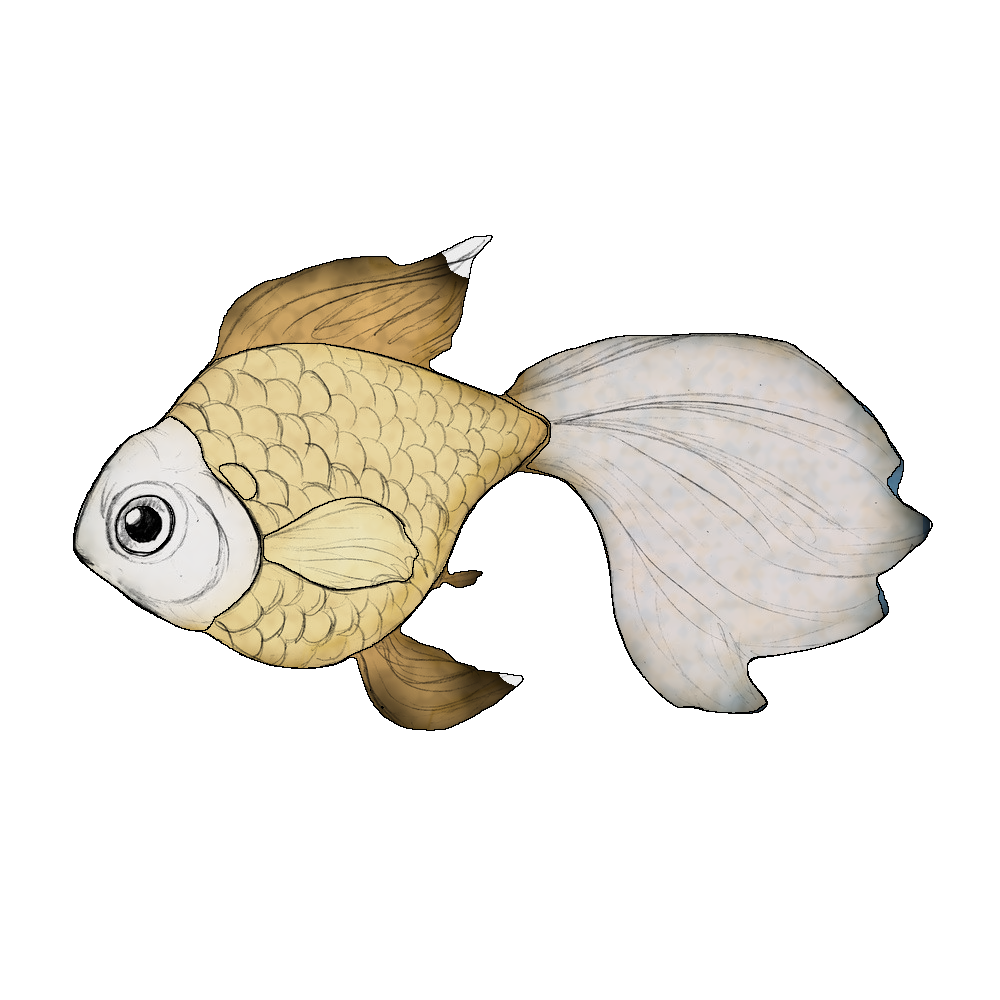}} \\

    \raisebox{-0.5\height}{\includegraphics[width=.25\linewidth]{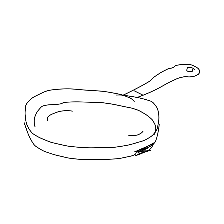}} &
    \raisebox{-0.5\height}{\includegraphics[width=.25\linewidth]{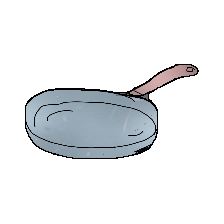}} &
    \raisebox{-0.5\height}{\includegraphics[width=.25\linewidth]{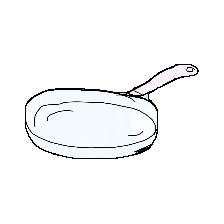}} &
    \raisebox{-0.5\height}{\includegraphics[width=.25\linewidth]{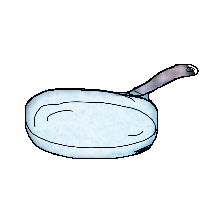}} \\

    \raisebox{-0.5\height}{\includegraphics[width=.25\linewidth]{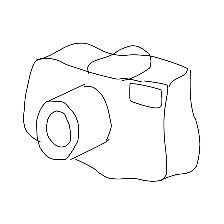}} &
    \raisebox{-0.5\height}{\includegraphics[width=.25\linewidth]{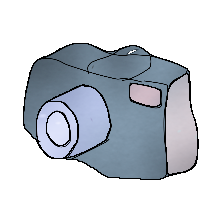}} &
    \raisebox{-0.5\height}{\includegraphics[width=.25\linewidth]{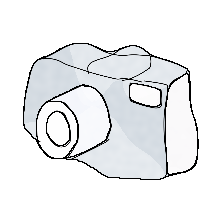}} &
    \raisebox{-0.5\height}{\includegraphics[width=.25\linewidth]{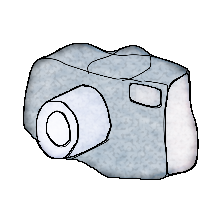}} \\

    \raisebox{-0.5\height}{\includegraphics[width=.25\linewidth]{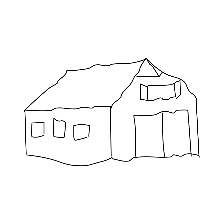}} &
    \raisebox{-0.5\height}{\includegraphics[width=.25\linewidth]{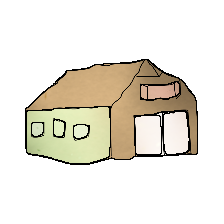}} &
    \raisebox{-0.5\height}{\includegraphics[width=.25\linewidth]{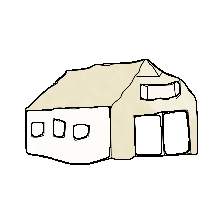}} &
    \raisebox{-0.5\height}{\includegraphics[width=.25\linewidth]{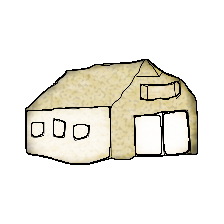}} \\

    \raisebox{-0.5\height}{\includegraphics[width=.25\linewidth]{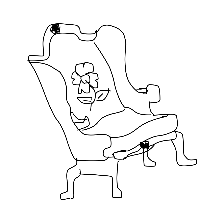}} &
    \raisebox{-0.5\height}{\includegraphics[width=.25\linewidth]{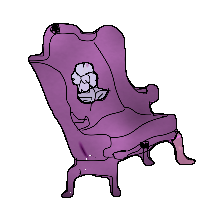}} &
    \raisebox{-0.5\height}{\includegraphics[width=.25\linewidth]{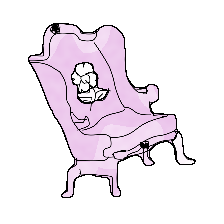}} &
    \raisebox{-0.5\height}{\includegraphics[width=.25\linewidth]{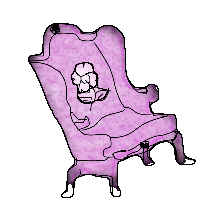}} \\
    
    \raisebox{-0.5\height}{\includegraphics[width=.25\linewidth]{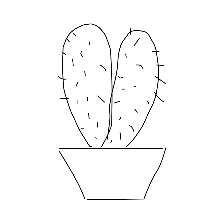}} &
    \raisebox{-0.5\height}{\includegraphics[width=.25\linewidth]{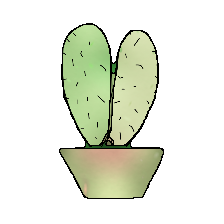}} &
    \raisebox{-0.5\height}{\includegraphics[width=.25\linewidth]{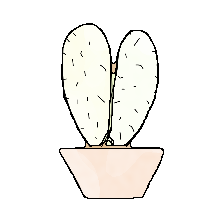}} &
    \raisebox{-0.5\height}{\includegraphics[width=.25\linewidth]{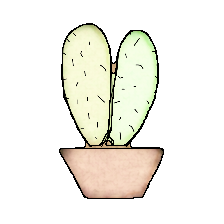}} \\
    
    \raisebox{-0.5\height}{\includegraphics[width=.25\linewidth]{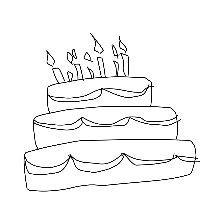}} &
    \raisebox{-0.5\height}{\includegraphics[width=.25\linewidth]{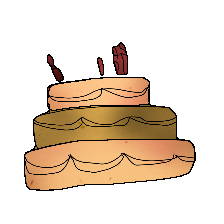}} &
    \raisebox{-0.5\height}{\includegraphics[width=.25\linewidth]{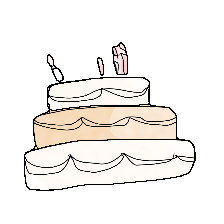}} &
    \raisebox{-0.5\height}{\includegraphics[width=.25\linewidth]{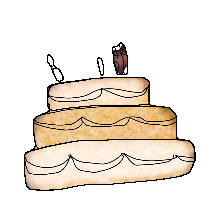}} \\
  \end{tabular}
  }
\caption{Examples demonstrating the style controllability of the proposed framework. Each row presents multiple stylized renderings of the same input sketch, obtained by varying procedural shader parameters while preserving the underlying segmentation and semantic color assignments.}
\label{fig:style_variants}
\end{figure}

Unlike learning-based colorization methods, which typically generate a single appearance for a given input, the proposed procedural framework provides explicit control over the final artistic style. As illustrated in Fig.~\ref{fig:style_variants}, diverse stylized renderings  including crayon, cel-shading, and watercolor are obtained by varying shader parameters and stochastic rendering primitives while preserving the same sketch segmentation and semantic color assignments. The resulting images exhibit distinct textural and stylistic characteristics, demonstrating the flexibility of the proposed pipeline to generate multiple high-quality paired training samples from a single input sketch.
\paragraph{Cel-shading} To introduce the appearance of traditional cel-shading, we remove the influence of the Gabor noise field, as cel-shaded artwork is characterized by smooth regions of uniform color separated by discrete tonal transitions. Accordingly, the propagated shadow field is heavily quantized to serve as the primary source of lightness variation, producing distinct shading bands instead of continuous gradients. Subtle procedural effects, including sparse scratches and stochastic drawing primitives, are retained to introduce slight imperfections and prevent the rendering from appearing overly synthetic.
\paragraph{Watercolor} To mimic a watercolor style, we model the characteristic pigment bleeding and soft, irregular boundaries. The segment mask is spatially warped to allow colors to diffuse both inward and outward from the original region boundaries, producing the imperfect edge transitions commonly observed in watercolor media. The influence of the shadow and curvature fields is substantially reduced, as strong geometric shading cues are generally absent in watercolor artwork. Instead, the high-frequency Gabor noise is replaced with low-frequency, multi-scale noise, whose contribution to the lightness channel is amplified to emulate the gradual pooling and uneven deposition of watercolor pigments.

\section{Limitations and Future Directions}
Despite generating visually plausible and stylistically diverse colored sketches, the proposed framework has some limitations. First, its performance depends heavily on the quality of the initial sketch segmentation, as errors in region extraction can propagate to semantic color assignment and subsequent procedural rendering. Second, the geometric descriptor fields are computed independently for each segmented region, limiting the ability to model fine-grained semantic relationships and maintain global object-level consistency. Lastly, Finally, although the procedural stylization pipeline provides explicit control over texture and artistic appearance, it relies on manually designed shaders and therefore cannot fully capture artist-specific styles or higher-level creative intent.

Several promising directions exist for future research. Incorporating more robust segmentation methods and hierarchical region representations could improve semantic consistency and reduce error propagation. The procedural shader library can be expanded with additional OSL-based rendering styles, such as oil painting with 3D effects, pencil rendering, and other non-photorealistic rendering techniques. Artist-specific procedural models or hybrid procedural–learning approaches could further enhance stylistic diversity by learning style parameters directly from reference artworks while preserving the controllability of the procedural framework. Finally, automatic style discovery could be explored to learn procedural style parameters directly from reference artworks, reducing the reliance on manually designed shaders while further increasing stylistic diversity and realism.

\section{Conclusion}
We presented TexSketch, a procedural framework for sketch colorization using semantic color assignment, geometric descriptor extraction, and Open Shading Language (OSL)-based procedural stylization. By combining contour-aware geometric features with programmable shader synthesis, our approach produces colored sketches that capture key characteristics of hand-drawn artwork, including contour emphasis, directional strokes, curvature-guided shading, and multi-scale procedural textures. Unlike manually curated datasets, the proposed framework is fully automated, scalable, and highly controllable, enabling the efficient generation of stylistically diverse paired sketch–color data for supervised learning. Experimental results demonstrate that the synthesized sketches are both perceptually plausible and stylistically consistent, while the procedural design offers explicit control over artistic appearance. TexSketch provides a valuable benchmark and data-generation framework for advancing sketch colorization and related sketch-understanding tasks beyond the limitations of existing anime-centric datasets.


\bibliography{main.bib}
\bibliographystyle{ieeetr}

\appendix
\section{Prompts}
\label{sec:prompts}

The prompts given to the vision-language model~\cite{Qwen-VL} are as follows.

\begin{lstlisting}[
  style=promptstyle,
  label=lst:prompt_qwen,
  caption={System prompt},
]
You are an expert anime, illustration, and abstract art colorist. You are handed a sketch where ONE specific region is highlighted in a vibrant bright red overlay with a dark red border.

Your goal is to color the image beautifully and dynamically, segment by segment.

CRITICAL PIPELINE RULES:
1. Look ONLY at the area covered by the bright red mask.
2. Identify the object. 
   - If concrete: Identify the exact semantic object (e.g., "hair", "skin", "jacket").
   - If abstract/surreal: Identify its compositional role (e.g., "foreground geometric shape", "flowing background ribbon", "negative space").
3. Determine the color strategy:
   - For concrete items: Use natural/logical colors (e.g., human skin tones, matching hair).
   - For abstract items: Establish a cohesive but highly varied conceptual color palette (e.g., neon cyberpunk, pastel vaporwave, vibrant primary colors). 
4. Review the history to maintain harmony, but strictly enforce contrast. If the image is abstract, YOU MUST NOT color distinct overlapping shapes with the same hue. Force yourself to pick a contrasting color from your palette.

Output format must be strictly raw JSON:
{
  "identified_object": "<what physical object or abstract shape is inside the red mask>",
  "palette_concept": "<e.g., 'earth tones', 'neon', 'concrete realistic'>",
  "H": <0-360>,
  "S": <0-100>,
  "L": <0-100>,
  "rationale": "<short justification explaining why this color provides good contrast or logical matching>"
}
\end{lstlisting}

\begin{lstlisting}[
  style=promptstyle,
  label=lst:prompt_qwen,
  caption={User prompt},
]
Analyze the image. One segment is highlighted with a red overlay.

--- ALREADY COLORED SECTOR HISTORY ---

{Previous HSL color choices and reasoning}

-------------------------------------

TASKS:
1. What is the object inside the red highlight? write it down in 'identified_object'.
2. Choose an appropriate HSL color. If it is skin or hair, use human-natural tones. If it matches a previous segment's item type, make it match or harmonize.
3. Return the JSON structure.
\end{lstlisting}

\end{document}